\documentclass[12pt]{article}
\pdfoutput=1
\usepackage{graphicx}
\usepackage{epsfig}
\usepackage{amsfonts}
\usepackage{amssymb}
\usepackage{amsmath}
\usepackage{dsfont}
\usepackage{pifont}
\usepackage{bbm}
\usepackage{multirow}
\usepackage{latexsym}
\usepackage{verbatim}
\usepackage{mcite}
\usepackage{cite}
\usepackage{units}
\usepackage[all]{xy}
\usepackage{cancel}
\usepackage{slashed}

\def\beq{\begin{equation}}
\def\eeq{\end{equation}}
\def\beqa{\begin{eqnarray}}
\def\eeqa{\end{eqnarray}}
\def\a{{\alpha}}
\def\b{{\beta}}
\def\g{{\gamma}}

\def\d{{\delta}}

\def\m{{\mu}}
\def\n{{\nu}}
\def\s{{\sigma}}

\def\dm{{\dot{\mu}}}

\def\bfone{\relax{\rm 1\kern-.35em 1}}

\def\hi{{\hat{i}}}
\def\hj{{\hat{j}}}
\def\hk{{\hat{k}}}
\def\hl{{\hat{l}}}

\newcommand{\cC}{{\cal C}}

\newcommand{\cN}{{\cal N}}

\newcommand{\cV}{{\cal V}}

\newcommand{\be}{\begin{equation}}
\newcommand{\ee}{\end{equation}}
\newcommand{\ben}{\begin{displaymath}}
\newcommand{\een}{\end{displaymath}}
\newcommand{\bea}{\begin{eqnarray}}
\newcommand{\eea}{\end{eqnarray}}

\newcommand{\bean}{\begin{eqnarray*}}
\newcommand{\eean}{\end{eqnarray*}}

\DeclareMathAlphabet{\mathpzc}{OT1}{pzc}{m}{it}
\begin{document}
\pagestyle{plain}
\makeatletter \@addtoreset{equation}{section} \makeatother
\renewcommand{\thesection}{\arabic{section}}
\renewcommand{\theequation}{\thesection.\arabic{equation}}
\renewcommand{\thefootnote}{\arabic{footnote}}
\setcounter{page}{1} \setcounter{footnote}{0}
\begin{titlepage}
\begin{flushright}
UUITP-01/14\\
\end{flushright}
\bigskip
\bigskip
\begin{center}
\vskip -5mm
{\LARGE \bf Lobotomy of Flux Compactifications} \\[6mm]%
\vskip 0.5cm
{\bf Giuseppe Dibitetto$^1$,\, Adolfo Guarino$^2$ \,and\, Diederik Roest$^3$}\let\thefootnote\relax\footnote{{\tt giuseppe.dibitetto@physics.uu.se, guarino@itp.unibe.ch, d.roest@rug.nl}}\\
\vskip 0.5cm
{\em
$^1$Institutionen f\"or fysik och astronomi, University of Uppsala, \\ Box 803, SE-751 08 Uppsala, Sweden \\[2mm]
$^2$ Albert Einstein Center for Fundamental Physics, Institute for Theoretical Physics, \\ Bern University, Sidlerstrasse 5, CH–3012 Bern, Switzerland \\[2mm]
$^3$ Centre for Theoretical Physics, University of Groningen, \\ Nijenborgh 4 9747 AG Groningen, The Netherlands \\}
\vskip 0.8cm
\end{center}
\vskip 1cm
\begin{center}
{\bf ABSTRACT}\\[3ex]
\begin{minipage}{13cm}
\small
We provide the dictionary between four-dimensional gauged supergravity and type II compactifications on $\mathbb{T}^6$ with metric and gauge fluxes in the absence of supersymmetry breaking sources, such as branes and orientifold planes. Secondly, we prove that there is a unique isotropic compactification allowing for critical points. It corresponds to a type IIA background given by a product of two  \mbox{$3$-tori} with $\textrm{SO}(3)$ twists and results in a unique theory (gauging) with a non-semisimple gauge algebra. Besides the known four AdS solutions surviving the orientifold projection to $\mathcal{N}=4$ induced by O$6$-planes, this theory contains a novel AdS solution that requires non-trivial orientifold-odd fluxes, hence being a genuine critical point of the $\mathcal{N}=8$ theory.
\end{minipage}
\end{center}
\vfill
\end{titlepage}


\tableofcontents

\newpage

\section{Introduction}
\label{sec:introduction}

In the last fifteen years a lot of work has been focused on the issue of moduli stabilisation in the context of string compactifications. Gauge fluxes and non-trivial internal geometries (referred to as metric fluxes in the simplest case of twisted tori) were proven to be needed for inducing a scalar potential to fix the moduli fields, at least perturbatively \cite{Dasgupta:1999ss,Gukov:1999ya}. In the renowned work of ref.~\cite{Maldacena:2000mw} fluxes were shown to always give rise to lower-dimensional theories with negative cosmological constant upon compactification. However, going beyond dimensional reductions on genuinely compact manifolds, one can circumvent the above no-go theorem and find no-scale Minkowski solutions by performing Scherk-Schwarz reductions on so-called flat group manifolds \cite{Scherk:1979zr}, like \emph{e.g.} $\textrm{ISO}(2) \times \textrm{ISO}(2)$. 

Another way of enlarging the set of possible lower-dimensional models is to add localised sources as extra ingredients, such as D-branes and O-planes. In particular, the presence of O-planes with negative tension turns out to be crucial in order to get a positive cosmological constant out of purely perturbative ingredients \cite{Giddings:2001yu,Danielsson:2009ff,Wrase:2010ew}. However the presence of localised sources has its own disadvantages like the explicit breaking of supersymmetry (with its potential instabilities \cite{Blaback:2012nf}), the possible failure of the supergravity approximation \cite{Bena:2012tx} or backreaction issues which have been pointed out and discussed in the literature (see \emph{e.g.} ref.~\cite{Bena:2012vz} and references therein). In this sense, compactifications of string theory \textit{without localised sources} turn out to be very robust as they preserve the maximal amount of supersymmetry, but are no longer appealing to find de Sitter universes or build brane models of Particle Physics. 

The prototype examples of lower-dimensional supergravities preserving maximal supersymmetry are the compactifications on $n$-spheres. Focusing on the case where no localised sources are present, the corresponding $\textrm{AdS}_{D-n} \times S^{n}$ solutions with SO($n+1$) gauge symmetry have been fairly explored in the literature. Such compactifications generically tend to suffer from the lack of scale separation, in the sense of not being proper lower-dimensional theories due to the fact that the KK scale and the AdS radius have comparable size \cite{Petrini:2013ika}.  At least in this case, one has a clue of the reason why this happens, \emph{i.e.} that maximal supersymmetry together with simple gauge groups with a rigid embedding constrain the theory too much to allow for the introduction of an extra scale. Still, despite this less appealing feature from a phenomenological viewpoint, such string backgrounds turn out to be relevant for holography, \emph{e.g.} type IIB on $\textrm{AdS}_{5} \times S^{5}$ or M-theory on $\textrm{AdS}_{4} \times S^{7}$.

Holographic applications increase the importance of the role of maximal gauged supergravities and the search for their AdS critical points. Due to the $S^7$ compactification of 11D supergravity, the $\textrm{SO}(8)$-gauged maximal supergravity in 4D \cite{deWit:1982ig} is of particular relevance and there has been a lot of progress in the analysis of its critical points. In this context, restricting oneself to smaller subsectors invariant under a given subgroup of the SO($8$) symmetry group has been a very fruitful approach to carry out a systematic search for critical points with non-trivial residual symmetry (see refs~\cite{Warner:1983du,Warner:1983vz} for cases with
$\textrm{SU}(3)$ and $\textrm{SO}(4)$ invariance). Later on, some new critical points with smaller \cite{Fischbacher:2010ec} and trivial \cite{Fischbacher:2009cj,Fischbacher:2011jx} residual symmetry were found, yielding the
first examples of stability without supersymmetry within a supergravity with such a high amount of supercharges.

The search for consistent gauged supergravities with extended supersymmetry has been boosted due to a new successful approach which is usually referred to as the embedding tensor formalism\footnote{See also refs~\cite{Nicolai:2000sc,Nicolai:2001sv} for previous results in three dimensions.} \cite{deWit:2002vt,deWit:2007mt}.
It is based on the idea of a duality-covariant formulation of gauged supergravities realised by promoting the corresponding deformation parameters to tensors w.r.t. the global symmetry group. This approach has led to substantial progress in classifying consistent gaugings of maximal supergravities \cite{deWit:2007mt} and has played a crucial role in finding the generalisation of the traditional $\textrm{SO}(8)$ theory with rigid embedding to a whole one-parameter family of theories \cite{DallAgata:2011aa,Dall'Agata:2012bb}. The physical relevance of this parameter in classifying inequivalent theories has been widely discussed \cite{Borghese:2012qm,Kodama:2012hu} and proven in the context of new $\textrm{SO}(8)$-gauged
maximal supergravity with $\textrm{SU}(3)$ residual symmetry, where the first examples of parameter-dependent mass spectra were found \cite{Borghese:2012zs} (see also refs~\cite{Dall'Agata:2012sx,Borghese:2013dja} for further analyses of critical points, refs~\cite{Guarino:2013gsa, Tarrio:2013qga} for domain-wall applications and refs~\cite{Anabalon:2013eaa,Lu:2014fpa} for black hole solutions).

The embedding tensor approach also turns out to be a valuable tool when linking extended gauged supergravities to flux compactifications \cite{deWit:2003hq}. The dictionary between fluxes in orientifolds of type II theories and embedding tensor deformations of half-maximal supergravities was worked out in refs~\cite{Dall'Agata:2009gv,Dibitetto:2010rg} and
subsequently used in ref.~\cite{Dibitetto:2011gm} to explore the set of critical points of $\cN=4$ compactifications of both type IIA with O$6$/D$6$ and type IIB with O$3$/D$3$. Since the set of AdS critical points found in the type IIA case turned out to be compatible with the total absence of localised sources, these were later interpreted as gauged $\cN=8$ supergravities in ref.~\cite{Dibitetto:2012ia}. These solutions became then novel examples of $\textrm{SO}(3)$-invariant critical points of maximal supergravity, one of which also happens to be non-supersymmetric and nevertheless tachyon-free.
The aim of the present paper is to extend the results of refs~\cite{Dibitetto:2011gm,Dibitetto:2012ia} by studying the most general backgrounds compatible with the absence of sources, thus containing
both orientifold-even and orientifold-odd fluxes. 

We will first derive the dictionary between type II fluxes and embedding tensor deformations in the \textbf{912} of $\textrm{E}_{7(7)}$. The derivation itself shows how geometric type II compactifications can be embedded in the much broader context of Exceptional Generalised Geometry (EGG) \cite{Grana:2009im,Aldazabal:2010ef,Berman:2010is,Berman:2011jh,Grana:2011nb}, one of the U-duality covariant frameworks that have been proposed for describing generalised string and M-theory backgrounds. We will briefly comment on other duality covariant approaches such as \emph{e.g.} Exceptional Field Theory (EFT) \cite{Aldazabal:2013mya,Aldazabal:2013via,Hohm:2013vpa,Hohm:2013uia}.  Keeping also duality covariance as the guiding principle, there have been some recent developments in the understanding of generalised Scherk-Schwarz reductions \cite{Berman:2012uy,Godazgar:2013dma,Godazgar:2013pfa,Godazgar:2013oba,Godazgar:2014sla}. These proposals, together with our present analysis, point towards a democratic formulation of fundamental ten- and eleven-dimensional degrees of freedom (d.o.f) as a good candidate to provide a higher-dimensional interpretation of the embedding tensor. A full-fledged reduction of the democratic (formulation of) type II supergravities \cite{Bergshoeff:2001pv}, supplemented by an \textit{appropriate} physical section condition to remove unphysical degrees of freedom in the lower-dimensional theory, goes well beyond the scope of this work.

Equipped with the aforementioned dictionary between type II fluxes and embedding tensor deformations, we will study the full set of $\textrm{SO}(3)$-invariant critical points compatible with geometric flux backgrounds on an isotropic $\mathbb{T}^{6}/(\mathbb{Z}_{2} \times \mathbb{Z}_{2})$ orbifold compactification of both IIA and IIB strings.  Remarkably, there turns out to exist a unique theory with specific IIA geometric fluxes allowing for such critical points. It has a non-semisimple gauge group arising from an $\,\textrm{SO}(3) \times \textrm{SO}(3)\,$ twisted torus reduction, and can be seen as the Scherk-Schwarz analogon of the $S^7$ compactification and the $\textrm{SO}(8)$ gauge group.
From a stringy perspective, the search for new compactifications without localised sources was motivated by the possibility of avoiding the issues which are typically introduced by O-planes when trying to reconcile the suppression of all corrections and large flux quanta together with tadpole cancellation \cite{Blaback:2013fca}. From a supergravity viewpoint, a complementary motivation is that of enriching the known classification of critical points of $\cN=8$ supergravity with $\textrm{SO}(3)$ residual symmetry by providing increasingly more new examples.

The paper is organised as follows. In Section~\ref{sec:N=8Sugra}, we first review the embedding tensor formalism applied to maximal supergravities in four dimensions and subsequently, we give the relation between its $\textrm{SL}(2) \times \textrm{SO}(6,6)$ formulation \cite{Dibitetto:2012ia} (naturally linked to fluxes) and its $\textrm{SU}(8)$ formulation \cite{LeDiffon:2011wt} (naturally related to fermion mass terms and scalar dynamics). In Section~\ref{sec:Fluxes/ET}, we start decomposing fields and deformations of maximal supergravity, which are arranged into irrep's of $\textrm{E}_{7(7)}$, with respect to the $\textrm{SL}(6)$ subgroup of diffeomorphisms under which the six internal coordinates transform as a vector. This will allow us to explicitly write down the dictionary embedding tensor/fluxes both in type IIB without O$3$-planes and in type IIA without O$6$-planes. We will follow the philosophy presented in ref.~\cite{Aldazabal:2010ef}, but restrict ourselves to geometric fluxes, \emph{i.e.} those deformations which have a clear higher-dimensional origin. In Section~\ref{sec:Vacua}, we will make use of the dictionary derived in the previous section in order to exhaustively study the set of critical points both in type IIB and in type IIA isotropic flux models without localised sources. While type IIB compactifications do not have new critical points, type IIA compactifications will turn out to have a new unstable AdS solution. Finally, some technical material is collected in the appendices~\ref{App:polyforms} and \ref{App:Dim_Red}.

\section{Gauged maximal supergravities in $D=4$}
\label{sec:N=8Sugra}

Maximal supergravity in four dimensions \cite{Cremmer:1979up,deWit:1982ig}, in its ungauged version, can be obtained from $\mathbb{T}^6$ reductions of type II supergravities in ten dimensions \cite{Cremmer:1997ct}. It enjoys an $\textrm{E}_{7(7)}$ global symmetry and its vectors (28 electric and 28 magnetic \cite{deWit:2002vt}) span the \textbf{56} representation. The bosonic sector of the theory also constains the metric field and $70$ scalar (physical) degrees of freedom parameterising an $\textrm{E}_{7(7)}/\textrm{SU}(8)$ coset element. In order to analyse the possible deformations (a.k.a. \textit{gaugings}) of maximal supergravity in a $\textrm{E}_{7(7)}$ covariant manner, the framework of the embedding tensor has been developed \cite{deWit:2007mt} and very successfully applied henceforth.

\subsection{Embedding tensor deformations : even \textit{vs} odd}

$\mathcal{N}=8$ ungauged supergravity can be deformed by promoting part of its $\textrm{E}_{7(7)}$ global symmetry to a gauge symmetry,  namely, by applying a gauging. A consistent gauging is completely specified by an embedding tensor transforming in the \textbf{912} of $\textrm{E}_{7(7)}$ denoted by ${\Theta_{\mathbb{M}}}^{\mathpzc{A}}$, where $\mathbb{M}=1,...,56$ and ${\mathpzc{A}=1,...,133}$ are a fundamental and an adjoint index respectively. This object selects which subset of the $\textrm{E}_{7(7)}$ generators $\left\lbrace t_{\mathpzc{A}=1,...,133} \right\rbrace$ become gauge symmetries after the gauging procedure. This is carried out through a covariant derivative $\nabla \rightarrow \nabla -g\, V^{\mathbb{M}} \,{\Theta_{\mathbb{M}}}^{\mathpzc{A}} \, t_{\mathpzc{A}} $, where $V^{\mathbb{M}}$ denote the vectors of the theory. As a consequence of the gauging, a non-Abelian gauge algebra
\be
\label{gauge_algebra}
\left[ X_{\mathbb{M}} , X_{\mathbb{N}} \right] = - {X_{\mathbb{M} \mathbb{N}}}^{\mathbb{P}} \, X_{\mathbb{P}} \hspace{15mm} \textrm{with} \hspace{15mm} {X_{\mathbb{M} \mathbb{N}}}^{\mathbb{P}} =
{\Theta_{\mathbb{M}}}^{\mathpzc{A}} \, {[t_{\mathpzc{A}}]_{\mathbb{N}}}^{\mathbb{P}} \ ,
\ee
is realised by the generators $X_{\mathbb{M}}$. By using the $\textrm{Sp}(56,\mathbb{R})$ invariant metric $\Omega_{\mathbb{M}\mathbb{N}}$ in the SouthWest-NorthEast (SW-NE) convention, one can define
$X_{\mathbb{M} \mathbb{N} \mathbb{P}} \equiv -{X_{\mathbb{M} \mathbb{N}}}^{\mathbb{Q}} \, \Omega_{\mathbb{Q} \mathbb{P}}$.
The embedding tensor in this form of generalised structure constants $X_{\mathbb{M} \mathbb{N} \mathbb{P}}$, needs then to satisfy the following set of quadratic constraints (QC) \cite{deWit:2007mt}
\be
\label{quadratic_const}
\Omega^{\mathbb{R} \mathbb{S}} \, X_{\mathbb{R} \mathbb{M} \mathbb{N}} \, X_{\mathbb{S} \mathbb{P} \mathbb{Q}}=0 \ ,
\ee
which guarantee the closure of the gauge algebra.

Describing the embedding tensor $X_{\mathbb{M} \mathbb{N} \mathbb{P}}$ as an $\textrm{E}_{7(7)}$ object is not very convenient in order to establish a neat correspondence between deformation parameters in supergravity and background fluxes in string theory. Instead, moving to an $\textrm{SL}(2) \times \textrm{SO}(6,6) \times \mathbb{Z}_{2}$ description turns out to facilitate this task \cite{deWit:2003hq,deWit:2007mt,Dibitetto:2012ia}. The relevant branching rule for this reads
\beq
\label{912_branching}
\begin{array}{cccc}
\textrm{E}_{7(7)} & \rightarrow & \textrm{SL}(2) \times \textrm{SO}(6,6) \times \mathbb{Z}_{2} &    \\[2mm]
\textbf{912} & \rightarrow & ( \textbf{2},\textbf{220} )_{(+)} \, + \,  (\textbf{2},\textbf{12} )_{(+)} \, + \,  (\textbf{1},\textbf{352'} )_{(-)} \, + \,  (\textbf{3},\textbf{32} )_{(-)} \\[2mm]
X_{\mathbb{M} \mathbb{N} \mathbb{P}} & \rightarrow & f_{\alpha MNP} \,\,\,\,\, \oplus \,\,\,\,\,  \xi_{\alpha M} \,\,\, \,\,\oplus \,\,\,\,\,  F_{M\dot{\mu}} \,\,\,\,\, \oplus \,\,\,  \Xi_{\alpha \beta \mu}\\[2mm]
\end{array}
\eeq
where $\,\alpha = 1,2\,$ and $\,M=1,...12\,$ respectively denote SL(2) and SO(6,6) fundamental indices. The spinorial\footnote{Except in the fermionic Lagrangian (\ref{Fermi_Lagrangian}) involving the eight gravitini $\psi^{\mathcal{I}}_{\mu}\,$, the index $\mu$ will never refer to coordinates in 4D space-time.} index $\,\mu \,(\dot{\mu})=1,...,32\,$ refers to the (conjugate) Majorana-Weyl representation of SO(6,6). Notice that the embedding tensor pieces with only bosonic indices are parity-even with respect to the $\mathbb{Z}_{2}$ factor, whereas those carrying a spinorial index turn out to be parity-odd \cite{Dibitetto:2011eu}. This $\mathbb{Z}_{2}$ action will be later on identified with an orientifold $\Omega_{p} (-1)^{F_{L}} \sigma$ action in the string theory side. Finally, in order to fit the irrep's in (\ref{912_branching}),  the symmetry properties $\,f_{\alpha MNP} = f_{\alpha [MNP]}\,$ and $\,\Xi_{\alpha \beta \mu}=\Xi_{(\alpha \beta) \mu}\,$ must hold together with the condition (\ref{Fslash_condition}) below.

The complete dictionary between the $\mathbb{Z}_{2}$-even $(+)$ pieces $\,f_{\alpha MNP}\,$ and $\,\xi_{\alpha M}\,$ in (\ref{912_branching}) and type II background fluxes has been worked out in ref.~\cite{Dibitetto:2011gm} in the context of half-maximal supergravity. Later, using the explicit truncation from maximal to half-maximal supergravity in ref.~\cite{Dibitetto:2011eu}, these string backgrounds were interpreted as gauged maximal supergravities in the special case of the absence of localised sources \cite{Dibitetto:2012ia}. Here we are going to extend these results and analyse more general backgrounds also including the $\mathbb{Z}_{2}$-odd $(-)$ fluxes $F_{M\dot{\mu}}$ and $\Xi_{\alpha \beta \mu}$ in (\ref{912_branching}). 

In order to do so, we first need the decomposition of the fundamental index $\mathbb{M}$ of $\textrm{E}_{7(7)}$ under $\textrm{SL}(2) \times \textrm{SO}(6,6) \times \mathbb{Z}_{2}$. It reads $\,\mathbb{M} \rightarrow \a M \oplus \mu\,$ according to the decomposition $\,\textbf{56} \rightarrow (\textbf{2},\textbf{12})_{(+)} +(\textbf{1},\textbf{32})_{(-)}\,$. After this splitting \cite{Dibitetto:2012ia}, the embedding tensor $\,X_{\mathbb{MNP}}\,$ consists of \textit{bosonic} components
\beq
\label{Xbosonic}
\begin{array}{cclc}
X_{\a M \b N \g P} & = & - \, \epsilon_{\b \g} \, f_{\a MNP} - \, \epsilon_{\b \g} \, \eta_{M [N}\, \, \xi_{ \a P]} - \, \epsilon_{\a (\b} \, \xi_{\g) M} \, \eta_{NP}& , \\[4mm]
X_{\a M \m \n} & = & -\dfrac{1}{4} \, f_{\a MNP} \, \left[ \g^{NP} \right]_{\m \n} - \dfrac{1}{4} \, \xi_{\a N} \, \left[ {\g_{M}}^{N} \right]_{\m \n} & , \\[4mm]
X_{\m \a M \n} = X_{\m \n \a M} & = & \dfrac{1}{8} \, f_{\a MNP} \, \left[ \g^{NP} \right]_{\m \n} - \dfrac{1}{24} \, f_{\a NPQ} \, \left[ {\g_{M}}^{NPQ} \right]_{\m \n} \\[3mm]
& + & \dfrac{1}{8} \, \xi_{\a N} \, \left[ {\g_{M}}^{N} \right]_{\m \n} - \dfrac{1}{8} \, \xi_{\a M} \, \cC_{\m \n} & , \\[2mm]
\end{array}
\eeq
involving an even number of fermionic indices (hence $\mathbb{Z}_{2}$-even) and being sourced by $\,f_{\a MNP}\,$ and $\,\xi_{\a M}\,$, as well as \textit{fermionic} ones
\beq
\label{Xfermionic}
\begin{array}{cclc}
X_{\mu \nu \rho} & = & - \dfrac{1}{2} \, {F_{M}}_{\dot{\n}} \, {{[\gamma_{N}]}_{\mu}}^{\dot{\n}} \, \left[ \g^{MN} \right]_{\nu \rho} & , \\[4mm]
X_{\mu \a M \b N} & = & -2 \, \epsilon_{\a \b} \, F_{[M \dot{\n}} \, {\left[ \g_{N]} \right]_{\m}}^{\dot{\n}} \, - \, 2 \, \eta_{MN} \, \Xi_{\a \b \m} & , \\[4mm]
X_{\a M \m \b N} = X_{\a M \b N \m} & = & \epsilon_{\a \b} \, {[\gamma_{N}]_{\mu}}^{\dot{\n}} \, F_{M\dot{\n}} \, + \, \Xi_{\a \b \n} \, {\left[ \g_{MN} \right]^{\n}}_{\m} \, + \, \Xi_{\a \b \m} \, \eta_{MN} & ,
\end{array}
\eeq
involving and odd number of fermionic indices (hence $\mathbb{Z}_{2}$-odd) and being sourced by $\,F_{M \dot{\mu}}\,$ and $\,\Xi_{\a \b \mu}\,$. This embedding tensor automatically satisfies a set of linear constraints required by supersymmetry,
provided that \cite{Aldazabal:2010ef,Dibitetto:2012ia}
\be
\label{Fslash_condition}
\begin{array}{lclclc}
\slashed{F}^{\m} & \equiv & \left[\g^{M}\right]^{\m\dot{\n}}\,F_{M\dot{\n}} & = & 0 & ,
\end{array}
\ee
but is still restricted by the set of quadratic constraints in (\ref{quadratic_const}) coming from the consistency of the gauging. The set of components in (\ref{Xbosonic}) specifies how half-maximal supergravity is embedded inside maximal \cite{Dibitetto:2011eu}, whereas the remaining components in (\ref{Xfermionic}) represent the completion from half-maximal to maximal supergravity \cite{Dibitetto:2012ia}. 

We refer the reader to Appendix~B in ref.\cite{Dibitetto:2012ia} for a detailed presentation of the conventions we have adopted all over the paper: the invariant $\,\eta_{MN}$ metric, $\,\gamma_{M}$-matrices, $\,\gamma_{M_{1}...M_{p}}$-forms and charge conjugation matrix $\,\mathcal{C}_{\m \n}\,$  of $\,\textrm{SO}(6,6)\,$ as well as the $\,\textrm{Sp}(56,\mathbb{R})\,$ symplectic matrix $\,\Omega_{\mathbb{MN}}\,$ and the $\,\textrm{SL}(2)\,$ Levi-Civita tensor $\,\epsilon_{\a \b}\,$.

\subsection{$T$-tensor, fermion masses and scalar dynamics}
\label{sec:SU(8)_formulation}

The embedding tensor $X_{\mathbb{MNP}}$ can be dressed up with the scalar fields in the theory\footnote{Upon SU$(8)$ gauge-fixing, the number of physical scalars is reduced from $133$ to the usual $70$ scalars in the $\mathcal{N}=8$ supergravity multiplet.} -- they are encoded into ${{\cV}^{\mathbb{M}}}_{\underline{\mathbb{M}}}(\phi_{\mathpzc{A}})\, \in \textrm{E}_{7(7)}/\textrm{SU}(8)$ -- resulting in the so-called $T$-tensor \cite{deWit:2007mt}. This is related to the embedding tensor of the previous section via
\beq
\label{change_of_basis}
T_{\underline{\mathbb{M}}\underline{\mathbb{N}}\underline{\mathbb{P}}}\,=\,\frac{1}{2}\,{{\cV}^{\mathbb{M}}}_{\underline{\mathbb{M}}}(\phi_{\mathpzc{A}})\,{{\cV}^{\mathbb{N}}}_{\underline{\mathbb{N}}}(\phi_{\mathpzc{A}})\, {{\cV}^{\mathbb{P}}}_{\underline{\mathbb{P}}}(\phi_{\mathpzc{A}})\,X_{\mathbb{MNP}} \ .
\eeq
We have underlined the indices just to stress the fact that $\,T_{\underline{\mathbb{M}}\underline{\mathbb{N}}\underline{\mathbb{P}}}\,$ in (\ref{change_of_basis}) depends on the scalar fields. The explicit expression of $\,{{\cV}^{\mathbb{M}}}_{\underline{\mathbb{M}}}\,$ at the origin of the scalar field space, namely at $\,\phi_{\mathpzc{A}}=0\,$, was derived in ref.~\cite{Dibitetto:2012ia}.

The $T$-tensor can be further decomposed under the $\,\textrm{SU}(8)\,$ maximal compact subgroup of $\,\textrm{E}_{7(7)}$. For this purpose, we need the branching rule $\,\textbf{56} \rightarrow \textbf{28} + \overline{\textbf{28}}\,$ which amounts to the index splitting $\,_{\underline{\mathbb{M}}} \rightarrow (\, _{\mathcal{IJ}} \, , \, ^{\mathcal{IJ}} \, )\,$, with $\,\mathcal{IJ}=-\mathcal{JI}\,$. Using the pieces $\,{T^{\mathcal{IJ KL}}}_{\mathcal{MN}}\,$ and $\,{T_{\mathcal{IJ}}}^{\mathcal{KL MN}}$ it is possible to take contractions sitting in the $\textbf{36}$ and $\textbf{420}\,$, namely,
\beq
\label{tracing}
\begin{array}{lclc}
\mathcal{A}^{\mathcal{IJ}}\,=\,\dfrac{4}{21} \, {T^{\mathcal{IKJL}}}_{\mathcal{KL}} & \hspace{8mm} \textrm{and} \hspace{8mm} & {\mathcal{A}_{\mathcal{I}}}^{\mathcal{JKL}}\,=\,2 \, {T_{\mathcal{MI}}}^{\mathcal{MJKL}} & ,
\end{array}
\eeq
which are directly identified with the \textit{scalar dependent} mass terms for the gravitini $\psi^{\,\,\mathcal{I}}_{\mu}$ and the dilatini $\chi_{\mathcal{IJK}}$ in the four-dimensional Lagrangian \cite{deWit:2007mt}
\be
\label{Fermi_Lagrangian}
\begin{array}{cclclc}
e^{-1} \, g^{-1} \, \mathcal{L}_{\textrm{fermi}} &=& \frac{\sqrt{2}}{2} \, \,\mathcal{A}_{\mathcal{I} \mathcal{J}}(\phi_{\mathpzc{A}})\, \,\overline{\psi}^{\,\,\mathcal{I}}_{\mu} \, \gamma^{\mu \nu} \,\psi^{\,\,\mathcal{J}}_{\nu} &+&
\,\mathcal{A}^{\mathcal{IJK},\mathcal{LMN}}(\phi_{\mathpzc{A}}) \,\, \overline{\chi}_{\mathcal{IJK}} \, \chi_{\mathcal{LMN}} \\[2mm]
&+&  \frac{1}{6} \, \,{\mathcal{A}_{\mathcal{I}}}^{\mathcal{JKL}}(\phi_{\mathpzc{A}}) \,\,\overline{\psi}^{\,\,\mathcal{I}}_{\mu} \, \gamma^{\mu} \, \chi_{\mathcal{JKL}} &+& \textrm{ h.c. } & , \\[2mm]
\end{array}
\ee
where $\,\mathcal{A}^{\mathcal{IJK},\mathcal{LMN}} \equiv \frac{\sqrt{2}}{144} \, \epsilon^{\mathcal{IJKPQR[LM}} \, {\mathcal{A}^{\mathcal{N}]}}_{\mathcal{PQR}}\,$.  The fermion mass terms (\ref{tracing}) are the fundamental objects in the SU(8) covariant formulation of maximal supergravity \cite{deWit:2007mt,LeDiffon:2011wt}.

After applying a gauging, \textit{i.e.} $X_{\mathbb{MNP}} \neq 0$, the dynamics of the scalar fields is governed by a scalar potential
\be
\label{V_SU8}
g^{-2} \,V = -\frac{3}{4} \, |\mathcal{A}_1|^{2} + \frac{1}{24} \, |\mathcal{A}_2|^{2} \ ,
\ee
where $\,|\mathcal{A}_1|^{2}=\mathcal{A}_{\mathcal{IJ}} \, \, \mathcal{A}^{\mathcal{IJ}}\,$ and $\,|\mathcal{A}_2|^{2}={\mathcal{A}_{\mathcal{I}}}^{\mathcal{JKL}} \, {\mathcal{A}^{\mathcal{I}}}_{\mathcal{JKL}}\,$ are positive defined. If turning off the vector fields in the theory, maximally symmetric solutions are obtained by solving the equations of motion of the scalars \cite{LeDiffon:2011wt}
\beq
\label{scalars_eom}
\begin{array}{ccc}
\mathcal{C}_{\mathcal{IJKL}} \,+\, \dfrac{1}{24} \, \epsilon_{\mathcal{IJKLMNPQ}} \, \mathcal{C}^{\mathcal{MNPQ}} &=& 0 \ ,
\end{array}
\eeq
with $\,\mathcal{C}_{\mathcal{IJKL}}={\mathcal{A}^{\mathcal{M}}}_{[\mathcal{IJK}}\,\mathcal{A}_{\mathcal{L}]\mathcal{M}} \, +\, \frac{3}{4} \, {\mathcal{A}^{\mathcal{M}}}_{\mathcal{N}[\mathcal{IJ}} \, {\mathcal{A}^{\mathcal{N}}}_{\mathcal{KL}]\mathcal{M}}\,$. At these solutions, the mass matrix for the physical scalars reads \cite{LeDiffon:2011wt,Borghese:2011en}
\beq
\label{Mass-matrix}
\begin{array}{ccl}
g^{-2} \, {\left(\textrm{mass}^{2}\right)_{\mathcal{IJKL}}}^{\mathcal{MNPQ}}  & =  &  \delta_{\mathcal{IJKL}}^{\mathcal{MNPQ}} \, \left( \frac{5}{24} \, \mathcal{A}^{\mathcal{R}}{}_{\mathcal{STU}} \, \mathcal{A}_{\mathcal{R}}{}^{\mathcal{STU}} - \frac{1}{2} \, \mathcal{A}_{\mathcal{RS}} \, \mathcal{A}^{\mathcal{RS}} \right) \\[2mm]
& + & 6 \, \delta_{[\mathcal{IJ}}^{[\mathcal{MN}} \, \left( \mathcal{A}_{\mathcal{K}}{}^{\mathcal{RS} |\mathcal{P}} \, \mathcal{A}^{\mathcal{Q}]}{}_{\mathcal{L}]\mathcal{RS}} - \frac{1}{4} \, \mathcal{A}_{\mathcal{R}}{}^{\mathcal{S} |\mathcal{PQ}]} \, \mathcal{A}^{\mathcal{R}}{}_{\mathcal{S}|\mathcal{KL}]} \right) \\[2mm] 
&-& \frac{2}{3} \, A_{[\mathcal{I}}{}^{[\mathcal{MNP}} \, \mathcal{A}^{\mathcal{Q}]}{}_{\mathcal{JKL}]} \ ,
\end{array}
\eeq
whereas the vector masses are given by
\beq
\label{Mass-matrix_vectors}
\begin{array}{cclc}
g^{-2} \, {\left(\textrm{mass}^{2}\right)_{\mathcal{IJ}}}^{\mathcal{KL}}  & =  & -\frac{1}{6} \, {\mathcal{A}_{[\mathcal{I}}}^{\mathcal{NPQ}} \, \delta_{\mathcal{J}]}^{[\mathcal{K}} \, {\mathcal{A}^{\mathcal{L}]}}_{\mathcal{NPQ}}  + \frac{1}{2} \,  {\mathcal{A}_{[\mathcal{I}}}^{\mathcal{PQ}[\mathcal{K}} \, {\mathcal{A}^{\mathcal{L}]}}_{\mathcal{J}]\mathcal{PQ}} & , \\
g^{-2} \, \left(\textrm{mass}^{2}\right)_{\mathcal{IJKL}}  & =  & \frac{1}{36} \, {\mathcal{A}_{[\mathcal{I}}}^{\mathcal{PQR}} \,\epsilon_{\mathcal{J}]\mathcal{PQRMNS}[\mathcal{K}}  \, {\mathcal{A}_{\mathcal{L}]}}^{\mathcal{MNS}} & .
\end{array}
\eeq

One of the main achievements in this work will be to compute the fermion mass terms in (\ref{tracing}) as a function of the embedding tensor pieces in (\ref{912_branching}) at the particular point $\,\phi_{\mathpzc{A}}=0\,$.  This point in field space might be or might not be compatible with the scalar equations of motion in (\ref{scalars_eom}).  Later we will look for solutions of these equations and then we will recast the discussion about the applicability of the correspondence between fermion mass terms and embedding tensor pieces we are deriving next.

\subsection{Fermion masses $\&$ embedding tensor}

Now we obtain the correspondence between fermion mass terms in (\ref{tracing}) and embedding tensor pieces in (\ref{912_branching}). In order to present the results, we need to split the $\,\textrm{SU(8)}\,$ index $\,\mathcal{I} \,\rightarrow\, i \, \oplus \, \hat{i}\,$ with $\,i \,,\,\hat{i} = 1, ... , 4\,$ according to its $\,\textrm{SU}(4) \,\times\, \textrm{SU}(4) \subset{\textrm{SU}(8)}\,$ maximal subgroup\footnote{Under the $\,\mathbb{Z}_{2}\,$ element in (\ref{912_branching}) truncating from maximal to half-maximal supergravity, the index $\,i\,$ is $\,\mathbb{Z}_{2}$-even and labels the four gravitini which are kept in the $\,\cN=4\,$ theory, whereas $\,\hat{i}\,$ is $\,\mathbb{Z}_{2}$-odd and labels the extra gravitini which form the completion to the full $\,\cN=8\,$ theory \cite{Dibitetto:2012ia}.}. This subgroup is identified with the $\,\textrm{SO}(6)_{\textrm{time-like}} \,\times\, \textrm{SO}(6)_{\textrm{space-like}} \subset{\textrm{SO}(6,6)}\,$ inducing the additional branchings in \textit{Lorentzian} coordinates
\beq
\label{SU(4)xSU(4)_branchings}
\begin{array}{lcc}
\textrm{SO}(6,6) & \supset & \textrm{SO}(6) \,\times\, \textrm{SO}(6)\\[1mm]
\textbf{12} & \rightarrow & (\textbf{6},\textbf{1}) + (\textbf{1},\textbf{6})\\[1mm]
\textbf{32} & \rightarrow & (\textbf{4},\textbf{4}) + (\bar{\textbf{4}},\bar{\textbf{4}})\\[1mm]
\textbf{32'} & \rightarrow & (\textbf{4},\bar{\textbf{4}}) + (\bar{\textbf{4}},\textbf{4})\\[1mm]
\end{array}
\hspace{8mm} \textrm{$\Leftrightarrow$} \hspace{8mm}
\begin{array}{lcc}
\textrm{SO}(6,6) & \supset & \textrm{SO}(6) \,\times\, \textrm{SO}(6)\\[1mm]
M & \rightarrow & m \,\,\oplus\,\, a \\[1mm]
\mu & \rightarrow & _{i \hat{j}} \,\,\oplus\,\, ^{i \hat{j}} \\[1mm]
\dot{\mu} & \rightarrow & {_{i\,}}^{\,\hat{j}} \,\,\oplus\,\, {^{i\,}}_{\hat{j}}\\[1mm]
\end{array} \ .
\eeq
In what follows we give the expressions for the fermion mass terms as a function of the $\mathbb{Z}_{2}$-even pieces $\,f_{\a MNP}\,$ , $\,\xi_{\a M}\,$ and the $\mathbb{Z}_{2}$-odd pieces $\,F_{M \dm}\,$ , $\,\Xi_{\a\b \m}\,$ of the embedding tensor further decomposed under (\ref{SU(4)xSU(4)_branchings}).

\subsubsection*{The gravitini mass $\,\mathcal{A}^{\mathcal{IJ}}$}
\label{A1A2_components}

We start by presenting the gravitini mass matrix in (\ref{Fermi_Lagrangian}). It consists of the purely unhatted and hatted blocks
\beq
\label{A1even}
\begin{array}{cclc}
g \, \mathcal{A}^{ij} & = & \dfrac{1}{24\,\sqrt{2}} \, \epsilon^{\a \b} \, \left(L_{\a}\right)^{*} \, \left[G^{m}\right]^{ik}\,\left[G^{n}\right]_{kl}\,\left[G^{p}\right]^{lj} \,f_{\b mnp} & , \\[4mm]
g \, \mathcal{A}^{\hat{i}\hat{j}} & = & \dfrac{i}{24\,\sqrt{2}} \, \epsilon^{\a \b} \, L_{\a} \, \left[G^{a}\right]^{\hat{i}\hat{k}}\,\left[G^{b}\right]_{\hat{k}\hat{l}}\,\left[G^{c}\right]^{\hat{l}\hat{j}} \,f_{\b abc} & ,
\end{array}
\eeq
together with the mixed one
\beq
\label{A1odd}
\begin{array}{cccl}
g \, \mathcal{A}^{i\hat{j}} \, = \, g \, \mathcal{A}^{\hat{j}i} & = & \dfrac{(1-i)}{4}\,\left(\left[G^{m}\right]^{ik}\,{F_{m\,k}}^{\hat{j}} \, + \, \delta^{\a\b} \, {\Xi_{\a\b}}^{i\hat{j}}\right) & .
\end{array}
\eeq
In the above expressions, we have introduced an $\,\textrm{SL(2)}\,$ vielbein $\,L_{\a}=(i,1)\,$ and a set of time-like (anti-self-dual) $\,[G_{m}]^{ij}\,$ and space-like (self-dual) $\,[G_{a}]^{\hi \hj}\,$ 't Hooft symbols, where $\,m\,,\,a=1,..,6\,$ respectively denote time-like and space-like direction of $\,\textrm{SO}(6,6)\,$ in Lorentzian coordinates\footnote{We again refer the reader to appendices B and D of ref.~\cite{Dibitetto:2012ia} for a detailed derivation of $\,{{\cV}^{\mathbb{M}}}_{\underline{\mathbb{M}}}\,$ at $\,\phi_{\mathpzc{A}}=0\,$ and also for conventions regarding $\,\textrm{SO}(6)_{\textrm{time/space-like}}\,$ 't Hooft symbols.}. The blocks in (\ref{A1even}) survive a truncation to half-maximal supergravity \cite{Dibitetto:2012ia} (see footnote~$5$) and are sourced by bosonic components of the embedding tensor $\,f_{\a MNP}\,$ and $\,\xi_{\a M}\,$. Contrary to them, those in (\ref{A1odd}) do not survive and are sourced by fermionic embedding tensor components $\,\Xi_{\a \b \m}\,$ and $\,F_{M \dot{\m}}\,$. 

\subsubsection*{The gravitini-dilatini couplings $\,{\mathcal{A}_{\mathcal{I}}}^{\mathcal{JKL}}$}

Let us now present the relation between the gravitini-dilatini coupling in (\ref{Fermi_Lagrangian}) and the pieces of the embedding tensor. The set of components comprising an even number of unhatted (equivalently hatted) indices \cite{Dibitetto:2012ia} consists of
\beq
\label{A2even}
\hspace{-0.3mm}
\begin{array}{lcll}
g \, {\mathcal{A}_{i}}^{jkl} & = & \frac{-1}{24 \sqrt{2}} \, \epsilon^{\a \b} \, \left(L_{\a}\right)^{*} \, \epsilon^{jkl i'} \, \left(\left[G^{m}\right]_{i'k'}\,\left[G^{n}\right]^{k'l'}\,\left[G^{p}\right]_{l'i} \,f_{\b mnp} \, + \, 6 \, \left[G^{m}\right]_{i'i} \, \xi_{\b m}\right) & , \\[4mm]
g \, {\mathcal{A}_{\hat{i}}}^{\hat{j}\hat{k}\hat{l}} & = & \frac{i}{3 \sqrt{2}} \, \epsilon^{\a \b} \, L_{\a} \, \epsilon^{\hj \hk \hl \hi'} \, \left(\left[G^{a}\right]_{\hat{i}'\hat{k}'}\,\left[G^{b}\right]^{\hat{k}'\hat{l}'}\,\left[G^{c}\right]_{\hat{l}'\hat{i}} \, f_{\b abc}\, - \, 6 \, \left[G^{a}\right]_{\hat{i}'\hat{i}} \xi_{\b a}\right) & , \\[4mm]
g \, {\mathcal{A}_{i}}^{j\hat{k}\hat{l}} & = & \frac{-i}{8 \sqrt{2}} \, \epsilon^{\a \b} \, L_{\a} \, \left(\left[G^{a}\right]^{\hat{k}\hat{l}}\left[G^{n}\right]_{ik}\,\left[G^{p}\right]^{kj} \,f_{\b anp} \, + \, \d_{i}^{j}\,\left[G^{a}\right]^{\hat{k}\hat{l}}\,\xi_{\b a}\right) & , \\[4mm]
g \, {\mathcal{A}_{\hat{i}}}^{\hat{j}kl} & = & \frac{-1}{8 \sqrt{2}} \, \epsilon^{\a \b} \, \left(L_{\a}\right)^{*}\,\left(\left[G^{m}\right]^{kl}\,\left[G^{a}\right]_{\hat{i}\hat{k}}\,\left[G^{b}\right]^{\hat{k}\hat{j}}\,f_{\b mab}\, - \, \d_{\hat{i}}^{\hat{j}}\,\left[G^{m}\right]^{kl} \, \xi_{\b m}\right) & ,
\end{array}
\eeq
and involves the bosonic embedding tensor pieces $\,f_{\a MNP}\,$ and $\,\xi_{\a M}\,$, whereas components involving an odd number of unhatted/hatted indices are given by
\beq
\label{A2odd}
\begin{array}{ccll}
g \, {\mathcal{A}_{i}}^{jk\hat{l}} & = & \frac{(1-i)}{2} \, \left(\left[G^{m}\right]^{jk}\,{F_{m\,i}}^{\hat{l}} \, + \,\d_{i}^{[j}\,\left[G^{m}\right]^{k]k'}\,{F_{m\,k'}}^{\hat{l}} \, - \,\delta^{\a\b}\,\d_{i}^{[j}\,{\Xi_{\a\b}}^{k]\hat{l}}\right) & , \\[3mm]
g \, {\mathcal{A}_{\hat{i}}}^{jkl} & = & \frac{(1+i)}{2} \, \left(L^{\a}\right)^{*}\,\left(L^{\b}\right)^{*}\,\epsilon^{ijkl}\,\Xi_{\a\b\,i\hat{i}} & , \\[3mm]
g \, {\mathcal{A}_{\hat{i}}}^{\hat{j}\hat{k}l} & = & -\frac{(1-i)}{2} \, \left(\left[G^{a}\right]^{\hat{j}\hat{k}}\,{{F_{a}}^{l}}_{\hat{i}} \, + \,\d_{\hat{i}}^{[\hat{j}}\,\left[G^{a}\right]^{\hat{k}]\hat{k}'}\,{{F_{a}}^{l}}_{\hat{k}'} \, + \, \delta^{\a\b} \, {\Xi_{\a\b}}^{l[\hat{k}}\,\d_{\hat{i}}^{\hat{j}]}\right) & , \\[3mm]
g \, {\mathcal{A}_{i}}^{\hat{j}\hat{k}\hat{l}} & = & \frac{(1+i)}{2} \, L^{\a}\,L^{\b} \, \epsilon^{\hat{i}\hat{j}\hat{k}\hat{l}}\,\Xi_{\a\b\,i \hat{i}} & ,
\end{array}
\eeq
and depend on the fermionic embedding tensor pieces $\,\Xi_{\a \b \m}\,$ and $\,F_{M \dot{\m}}\,$. Notice that in the relation~\eqref{A1odd} we got rid of the space-like contraction $\,\left[G^{a}\right]^{\hat{j}\hat{k}}\,{{F_{a}}^{i}}_{\hat{k}}\,$ by
solving the linear constraint in \eqref{Fslash_condition}, which takes the following form when choosing $\,\textrm{SO}(6,6)\,$ Lorentzian coordinates
\beq
\left[G^{m}\right]^{ik}\,{F_{m\,k}}^{\hat{j}} \, + \, \left[G^{a}\right]^{\hat{j}\hat{k}}\,{{F_{a}}^{i}}_{\hat{k}} \, = \, 0 \ .
\eeq

The full mapping between the fermion mass terms $\,\{ \mathcal{A}^{\mathcal{IJ}} \, , \, {\mathcal{A}_{\mathcal{I}}}^{\mathcal{JKL}}\,\}$ and the embedding tensor pieces $\, \{f_{\alpha MNP}\, , \,\xi_{\alpha M}\, , \,F_{M \dot{\mu}}\, , \,\Xi_{\alpha \beta  \mu}\}$ in eqs~(\ref{A1even})-(\ref{A2odd}) represents one of the main results of the paper. Combining this mapping with the SU(8) formulation of maximal supergravity described in Section~\ref{sec:SU(8)_formulation}, we will be able to explore the scalar dynamics induced by generic configurations of the embedding tensor. However, in order to establish connections to type II string theory, we still need to derive the precise correspondence between type II background fluxes and embedding tensor components. This will be our goal in the next section.

\section{Gauged maximal supergravity from type II strings}
\label{sec:Fluxes/ET}

\begin{figure}[t!]
\begin{center}
\scalebox{0.22}[0.22]{
\includegraphics[keepaspectratio=true]{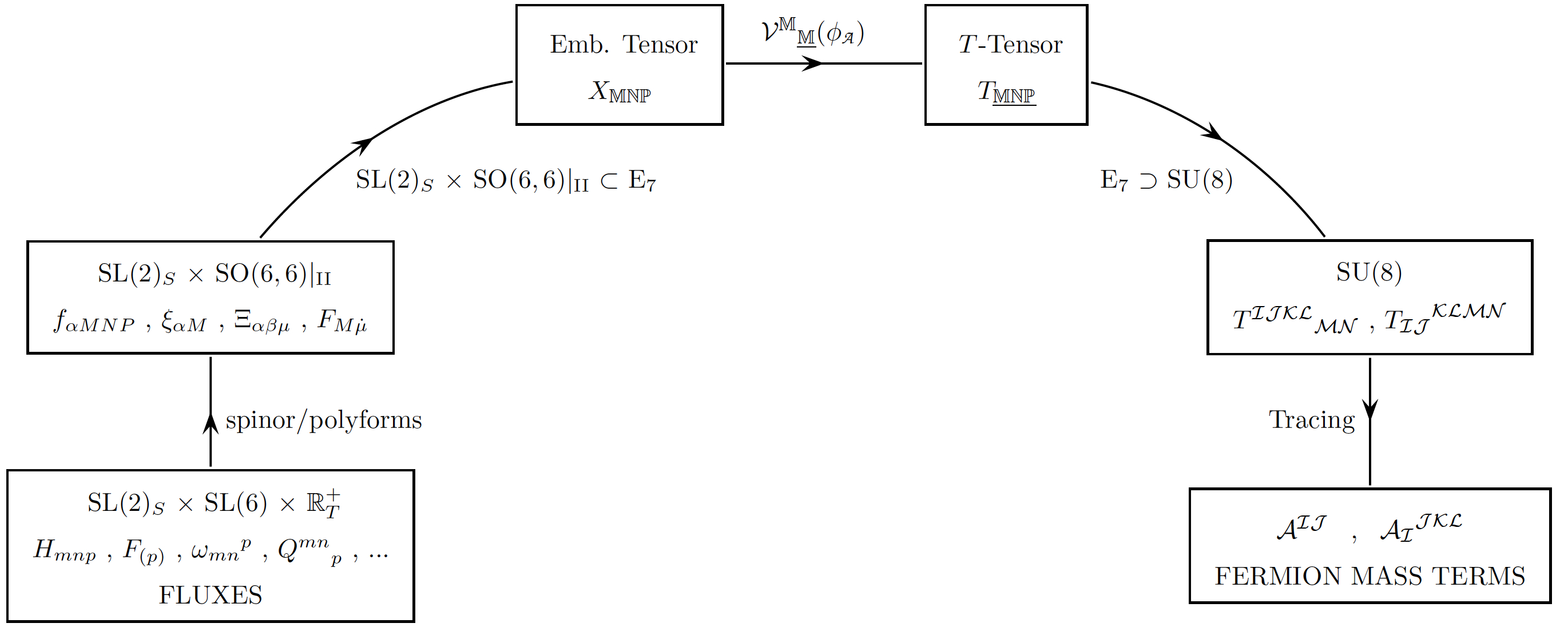}}
\end{center}
\vspace{-5mm}
\caption{{\it Diagram sketching the connection between type II flux backgrounds (lower-left) and fermion mass terms (lower-right) passing through the set of intermediate steps described in the main text.}}
\label{fig:diagram}
\end{figure}

In this section we discuss the correspondence between the ingredients in type II flux models and their related quantities on the supergravity side according to group theory. We will pay special attention to the dictionary between type II background fluxes and the embedding tensor, which has been found to totally encode the set of possible deformations of the free (ungauged) theory \cite{deWit:2007mt}. 

After finding the precise type II fluxes$\,\leftrightarrow\,$embedding tensor dictionary, we will be able to connect flux backgrounds to fermion mass terms (and thus to explore the scalar dynamics) following the path depicted in Figure~\ref{fig:diagram}. This procedure was introduced in ref.~\cite{Dibitetto:2012ia}, where the correspondence between fluxes and fermion masses was derived in the absence of fluxes related to spinorial components of the embedding tensor, \textit{i.e.} $\,F_{M \dot{\m}} \,=\, \Xi_{\a \b \m} \,=\, 0\,$. In this section we are extending those results by considering spinorial fluxes as well, hence completing the correspondence between fluxes and fermion masses. In particular, we would like to focus on geometric flux backgrounds\footnote{The full non-geometric dictionary with some applications will be presented in a companion paper \cite{non-geometric}.}. Hence we will add to the geometric type II backgrounds studied in ref.~\cite{Dibitetto:2012ia} only those spinorial fluxes which have a well-understood origin in string theory, like \emph{e.g.}, in type IIB, the R-R fluxes $\,F_{1}\,$ and $\,F_{5}\,$ or the metric flux $\,{\omega_{mn}}^{p}\,$ amongst others. The type II  fluxes/embedding tensor dictionary, together with the embedding tensor/fermion masses correspondence in eqs~(\ref{A1even})-(\ref{A2odd}), will be a valuable tool to explore moduli stabilisation in the last section of the paper.

\subsection{The type II embedding inside $\textrm{E}_{7(7)}$}
\label{sec:GroupI_embedding_II}

Maximal supergravities can be obtained from type II string compactifications preserving all the original supercharges \cite{Cremmer:1997ct}, \textit{e.g.} upon $\,\mathbb{T}^6\,$ toroidal compactifications (with coordinates $y^{m}$, $\,m=1,...,6$) from ten down to four dimensions (10D $\rightarrow$ 4D). The different fields living in the 4D theory organise into representations of the diffeomorphisms' group along the internal six-dimensional space, \textit{i.e.} $\textrm{SL}(6)\,$, which appears as (part of) a global symmetry of the 4D theory. However, some degeneracies between 4D fields occur at the level of their $\,\textrm{SL}(6)\,$ behaviour: as an example, there are several scalars which are singlets under $\textrm{SL}(6)$. This points towards a desirable enhancement of the global symmetry group in the lower-dimensional theory lifting the degeneracy between fields. Indeed, the 4D theory happens to enjoy a bigger global symmetry group:
the exceptional $\,\textrm{E}_{7(7)}\,$ group also known as the U-duality group \cite{Cremmer:1979up,Cremmer:1997ct}. In addition to the internal diffeomorphisms, it accounts for constant shifts of the gauge fields along the internal space
coordinates and also stringy transformations as T-duality or S-duality \cite{Hull:2007zu,Pacheco:2008ps,Aldazabal:2010ef,Coimbra:2011ky}. Since the lower-dimensional states are firstly labelled according to their behaviour under internal $\textrm{SL}(6)$ diffeomorphisms, the natural question is then how these are embedded inside the U-duality group. In the case of type II strings, the answer is given by the series of maximal subgroups \cite{deWit:2003hq}
\beq
\label{SL2xSL6_embedding}
\begin{array}{ccccc}
\textrm{E}_{7(7)} &\supset& \textrm{SL}(3) \times \textrm{SL}(6) &\supset& \textrm{SL}(2) \times \textrm{SL}(6) \times \mathbb{R}^{+} \ ,
\end{array}
\eeq
so additional $\textrm{SL}(2)$ and $\mathbb{R}^{+}$ labels can be used in order to unambiguously classify states in the lower-dimensional theory.
As a bi-product, the $\,\left. \textrm{SL}(2)_{S} \times \textrm{SO}(6,6) \right|_{\textrm{II}}\,$ embedding of maximal supergravity can be obtained by demanding the branching
\beq
\label{SL2xSO66_II_embedding}
\begin{array}{ccccc}
\textrm{E}_{7(7)} &\supset& \left. \textrm{SL}(2)_{S} \times \textrm{SO}(6,6) \right|_{\textrm{II}} &\supset& \textrm{SL}(2)_{S} \times \textrm{SL}(6) \times \mathbb{R}^{+}_{T} \ ,
\end{array}
\eeq
to produce the same decompositions as (\ref{SL2xSL6_embedding}). When applied to the relevant U-duality representations appearing in the $\,\textrm{E}_{7(7)}\,$ description of maximal supergravity,
\textit{i.e.} the \textbf{56} (vectors), \textbf{133} (scalars) and \textbf{912} (embedding tensor), one obtains the results displayed in Table~\ref{table:typeII_embedding}.
\begin{table}[t!]
\renewcommand{\arraystretch}{1.50}
\begin{center}
\scalebox{0.76}[0.80]{
\begin{tabular}{|ccccl|}
\hline
$\textrm{E}_{7(7)}$ & $\supset$ & $ \left.\textrm{SL}(2)_{S} \times \textrm{SO}(6,6)\right|_{\textrm{II}}\,$ & $\supset$ & $ \textrm{SL}(2)_{S} \times \textrm{SL}(6) \times \mathbb{R}^{+}_{T} $ \\[2mm]
\hline
\hline
$\textbf{56}$ & $\rightarrow$ & $\textbf{(2,12)}$ & $\rightarrow$ & $\textbf{(2,6)}_{(+\frac{1}{2})} + \textbf{(2,6')}_{(-\frac{1}{2})}$ \\[2mm]
& & $\textbf{(1,32)}$ & $\rightarrow$ & $\textbf{(1,6')}_{(+1)} + \textbf{(1,20)}_{(0)} + \textbf{(1,6)}_{(-1)}$ \\[2mm]
\hline
$\textbf{133}$ & $\rightarrow$ & $\textbf{(1,66)}$ & $\rightarrow$ & $\textbf{(1,15)}_{(+1)} + \textbf{(1,1+35)}_{(0)} + \textbf{(1,15')}_{(-1)}$\\[2mm]
& & $\textbf{(3,1)}$ & $\rightarrow$ & $\textbf{(3,1)}_{(0)}$ \\[2mm]
& & $\textbf{(2,32')}$ & $\rightarrow$ & $\textbf{(2,1)}_{(+\frac{3}{2})} + \textbf{(2,15')}_{(+\frac{1}{2})} + \textbf{(2,15)}_{(-\frac{1}{2})} + \textbf{(2,1)}_{(-\frac{3}{2})}$ \\[2mm]
\hline
$\textbf{912}$ & $\rightarrow$ & $\textbf{(2,12)}$ & $\rightarrow$ & $\textbf{(2,6)}_{(+\frac{1}{2})} + \textbf{(2,6')}_{(-\frac{1}{2})}$\\[2mm]
& & $\textbf{(2,220)}$ & $\rightarrow$ & $\textbf{(2,20)}_{(+\frac{3}{2})} + \textbf{(2,6+84)}_{(+\frac{1}{2})} + \textbf{(2,\, 6'+84')}_{(-\frac{1}{2})} + \textbf{(2,20)}_{(-\frac{3}{2})}$\\[2mm]
& & $\textbf{(3,32)}$ & $\rightarrow$ & $\textbf{(3,6')}_{(+1)} + \textbf{(3,20)}_{(0)} + \textbf{(3,6)}_{(-1)}$ \\[2mm]
& & $\textbf{(1,352')}$ & $\rightarrow$ & $\textbf{(1,6)}_{(+2)} + \textbf{(1,6'+84')}_{(+1)} + \textbf{(1,70+20+70')}_{(0)} + \textbf{(1,6+84)}_{(-1)} + \textbf{(1,6')}_{(-2)}$ \\[2mm]
\hline
\end{tabular}
}
\end{center}
\caption{{\it Branching of $\,\textrm{E}_{7(7)}\,$ representations according to the type II group theoretical embedding of maximal supergravity.}}
\label{table:typeII_embedding}
\end{table}

\subsection{O-planes and orientifolds}

As briefly mentioned in the introduction, the inclusion of O-planes in the string compactification scheme breaks supersymmetry explicitly \cite{Giddings:2001yu,Grana:2005jc}. In addition, having O-planes as localised sources also induces orientifold actions which are the combination of three $\mathbb{Z}_{2}$ gradings: two of them act at the level of the worldsheet fields whereas the last one acts at the level of target space coordinates. 

The worldsheet orientifold action is a combination of the so-called fermion number $(-1)^{F_{L}}$ in the left-moving sector and the worldsheet parity $\Omega_{p}$ which acts on the corresponding fields by exchanging left- and right-movers. Under the combined $\,(-1)^{F_{L}}\Omega_{p}\,$ action, the type II fields $g \,,\, \phi \,,\, C_{0} \,,\, C_{(3)}$ and $C_{(4)}$ are parity-even whereas $B_{(2)} \,,\, C_{(1)}$ and $C_{(2)}$ are parity-odd. The target space orientifold involution $\sigma$, instead assigns positive parity to the coordinates along the O-plane worldvolume and a negative one to the transverse coordinates \cite{Grana:2005jc}. We will describe in detail the O3-plane ($\sigma_{\textrm{O3}}$) and O6-plane ($\sigma_{\textrm{O6}}$) orientifold involutions in the next sections.

The ultimate aim of this work is to remove orientifolds in type II flux compactifications. Unorientifolding type II compactifications means to place the different fluxes and fields inside bosonic or spinorial irrep's of $\textrm{SO}(6,6)$ according to whether they are allowed (\mbox{$\mathbb{Z}_{2}$-even}) or forbidden ($\mathbb{Z}_{2}$-odd) by the orientifold action $(-1)^{F_{L}}\,\Omega_{p}\,\sigma$.

\subsection{Unorientifolding type IIB with O3-planes}
\label{sec:GroupI_embedding_IIB}

Type IIB backgrounds with O$3$-planes (and the corresponding D$3$-branes) are characterised by supersymmetry-breaking extended sources which are completely localised in the six-dimensional internal space.
Their position can be chosen as
\beq
\begin{array}{lcccc}
\textrm{O$3$-plane} \, : & & & \underbrace{\times \, \vert \, \times \, \times \, \times}_{D = 4} \, \, \underbrace{- \, - \, -\, - \, - \, -}_{m} & 
\end{array}
\notag
\eeq
where $m$ spans the fundamental representation of $\textrm{SL}(6)$. The orientifold involution is in this case defined by
\beq
\label{sigma_O3}
\sigma_{\textrm{O}3} \,\,: \,\, (\, y^{1} \, , \, y^{2} \, , \, y^{3} \, , \, y^{4} \, , \, y^{5} \, , \,y^{6} \, ) 
\,\, \rightarrow \,\, 
(\, -y^{1} \, , \, -y^{2} \, , \, -y^{3} \, , \, -y^{4} \, , \, -y^{5} \, , \, -y^{6} \, ) \ .
\eeq
We immediately predict that the IIB fluxes/embedding tensor dictionary in this case will be $\textrm{SL}(6)$-covariant since the $\sigma_{\textrm{O}3}$ orientifold involution (\ref{sigma_O3}) treats all the internal coordinates on equal footing. Indeed, by taking a look into Table~\ref{table:typeII_embedding}, one observes that it is \textit{completely democratic with respect to 6D Hodge duality} along the internal space. Equivalently, in terms of the content of $\textrm{SL}(6)$ states, whenever there is a 0-form state then also a 6-form appears and the same with pairs of (1,5)-forms and (2,4)-forms. Thus, in order to obtain the IIB dictionary, one needs to decompose fields and deformations of maximal supergravity (which naturally group into $\textrm{E}_{7(7)}$ irrep's) into states labelled by their behaviour with respect to diffeomorphisms, \emph{i.e.} $\textrm{SL}(6)$ and their $ST$ weights
\be
\label{branching_IIB_S}
\begin{array}{ccc}
\textrm{SL}(2)_{S} \times \textrm{SL}(6) \times \mathbb{R}^{+}_{T} & \supset & \textrm{SL}(6) \times \mathbb{R}^{+}_{S} \times \mathbb{R}^{+}_{T} \ .
\end{array}
\ee
Some relevant $\textrm{SL}(2)_{S}  \rightarrow \mathbb{R}^{+}_{S}$ branchings are $\textbf{2} \rightarrow \textbf{1}_{(-1/2)} + \textbf{1}_{(1/2)}$ and $\textbf{3} \rightarrow \textbf{1}_{(-1)} + \textbf{1}_{0} + \textbf{1}_{(1)}$. 
The above decomposition in (\ref{branching_IIB_S}) will be carried out for the $\,\textbf{56}\,$, $\,\textbf{133}\,$ and $\,\textbf{912}\,$ of $\textrm{E}_{7(7)}$, which respectively describe vectors, scalars and deformations of maximal
supergravity.
\\

\begin{table}[t!]
\begin{center}
\scalebox{0.80}[0.85]{
\begin{tabular}{|c||c|c|c|c|}
\hline
B/F & $\sigma_{\textrm{O}3}$ & $(-1)^{F_{L}}\,\Omega_{p}$ & operator & $\textrm{SL}(6) \times \mathbb{R}^{+}_{S} \times \mathbb{R}^{+}_{T}$ \\[1mm]
\hline\hline
F & $-$ & $+$ & $\partial_{m}$ & $\textbf{6'}_{(0;+1)}$ \\[2mm]
\hline
\end{tabular}
}
\end{center}
\caption{{\it The physical internal derivatives in type IIB compactifications. It is the combination $\,(-1)^{F_{L}}\,\Omega_{p}\,\sigma_{\textrm{O3}}\,$ of fermionic number, worldsheet parity and orientifold involution what determines that $\partial_{m}$ is completely projected out by the presence of O3-planes. As a consequence, all its components sit inside a fermionic (F) irrep of $\textrm{SO}(6,6)$.}}
\label{table:IIB_Der}
\end{table}

\noindent \textbf{The \textbf{56} representation :} From the U-duality point of view, the $\,\textbf{56}\,$ representation can be used to introduce a $\textrm{E}_{7(7)}$-derivative $\,\partial_{\mathbb{M}}\,$ defining an infinitesimal $\textrm{E}_{7(7)}$-variation in the U-duality space \cite{Aldazabal:2010ef}. Following the upper decomposition in Table \ref{table:typeII_embedding}, and further performing the branching described in \eqref{branching_IIB_S}, one can identify
the \textit{physical} derivatives $\,\partial_{m}\equiv \partial/\partial y^{m}$ related to $\textrm{SL}(6)$ variations. This identification relies on the singlet nature of the internal coordinates under type IIB S-duality (vanishing $\mathbb{R}^{+}_{S}$ charge). 
Moreover note that, since the operator $\partial_{m}$ is not constructed out of string oscillators, it is naturally even under the worldsheet orientifold action.
The result is described in Table~\ref{table:IIB_Der}.\\

\begin{table}[t!]
\begin{center}
\scalebox{0.80}[0.85]{
\begin{tabular}{|c||c|c|c|c|}
\hline
B/F & $\sigma_{\textrm{O}3}$ & $(-1)^{F_{L}}\,\Omega_{p}$ & IIB field & $\textrm{SL}(6) \times \mathbb{R}^{+}_{S} \times \mathbb{R}^{+}_{T}$ \\[1mm]
\hline\hline
\multirow{6}{*}{B} & $+$ & $+$ & $\phi$ & $\textbf{1}_{(0;\,0)}$ \\[2mm]
\cline{2-5} & $+$ & $+$ & ${e_{m}}^{n}$ & $\textbf{35}_{(0;\,0)}$ \\[2mm]
\cline{2-5} & $+$ & $+$ & ${e_{m}}^{m}$ & $\textbf{1}_{(0;\,0)}$ \\[2mm]
\cline{2-5} & $+$ & $+$ & $C_{0}$ & $\textbf{1}_{(+1;\,0)}$ \\[2mm]
\cline{2-5} & $+$ & $+$ & $C_{mnpq}$ & $\textbf{15}_{(0;+1)}$ \\[2mm]
\hline\hline
\multirow{5}{*}{F} & $+$ & $-$ & $B_{mn}$ & $\textbf{15'}_{(-\frac{1}{2};+\frac{1}{2})}$ \\[2mm]
\cline{2-5} & $+$ & $-$ & $B_{mnpqrs}$ & $\textbf{1}_{(-\frac{1}{2};-\frac{3}{2})}$ \\[2mm]
\cline{2-5} & $+$ & $-$ & $C_{mn}$ & $\textbf{15'}_{(+\frac{1}{2};+\frac{1}{2})}$ \\[2mm]
\cline{2-5} & $+$ & $-$ & $C_{mnpqrs}$ & $\textbf{1}_{(+\frac{1}{2};-\frac{3}{2})}$ \\[2mm]
\hline
\end{tabular}
}
\end{center}
\caption{{\it The physical scalars from type IIB compactifications mapped into states in the $\textbf{133}$ of $\textrm{E}_{7(7)}$. Note that it is the combination $(-1)^{F_{L}}\,\Omega_{p}\,\sigma_{\textrm{O}3}$ of fermionic number, worldsheet parity and orientifold involution what determines which states are bosonic (B) and fermionic (F). It is worth mentioning that, in order to get the correct number of physical degrees of freedom (\emph{i.e.}
$70 = 38_{\textrm{B}} + 32_{\textrm{F}}$), one needs to subtract the compact directions inside the vielbein. }}
\label{table:IIB_Gauge}
\end{table}

\noindent \textbf{The \textbf{133} representation :} This representation of the U-duality group accommodates scalar fields $\,\phi_{\mathpzc{A}}\,$, with $\mathpzc{A}=1,...,133\,$, associated to the generators of the $\textrm{E}_{7(7)}$ duality group of maximal supergravity. These scalars, carrying the $ \textrm{SL}(2)_{S} \times \textrm{SL}(6) \times \mathbb{R}^{+}_{T} $ charges displayed in Table~\ref{table:typeII_embedding}, precisely match the dimensional reduction of the democratic 10D fields in type IIB supergravity \cite{Bergshoeff:2001pv} when keeping pure scalars, \textit{i.e.} components with no legs along the 4D spacetime, as well as two-forms, \textit{i.e.} components with two legs dual to scalars upon 4D Hodge duality\footnote{It would be very interesting to understand the relation between this set of two-forms and the $(\beta,\gamma)$-fields introduced in ref.~\cite{Aldazabal:2010ef}.}. Upon local SU$(8)$ gauge fixing, the physical scalars -- which carry $70$ degrees of freedom in total -- can be \textit{aligned} with the pure scalars in the above reduction\footnote{In this work we are not considering non-geometric setups where the remaining  $63$ fields have a topologically non-trivial flux \cite{Aldazabal:2010ef}.}. These $70$ scalars split up into $38$ orientifold-allowed ones arising from
\be
\label{IIB_B_scalars}
\big\lbrace \underbrace{\hspace{2mm} \phi \hspace{4mm} , \hspace{4mm} {e_{m}}^{n} \hspace{4mm} , \hspace{4mm} {e_{m}}^{m} \equiv \textrm{Tr}(e)\hspace{2mm}}_{\textrm{NS-NS}} \hspace{4mm} ,
\hspace{4mm} \underbrace{\hspace{2mm} C_{0} \hspace{4mm} , \hspace{4mm} C_{mnpq}\hspace{2mm}}_{\textrm{R-R}} \big\rbrace \ ,
\notag
\ee
where the correct counting is reproduced upon subtracting the $15$ compact $\textrm{SO}(6)$ directions inside ${e_{m}}^{n}$, and $32$ orientifold-forbidden ones coming from
\be
\label{IIB_F_scalars}
\big\lbrace \underbrace{\hspace{2mm} B_{mn} \hspace{4mm} , \hspace{4mm} B_{mnpqrs} \hspace{2mm}}_{\textrm{NS-NS}} \hspace{4mm} ,
\hspace{4mm} \underbrace{\hspace{2mm} C_{mn} \hspace{4mm} , \hspace{4mm} C_{mnpqrs} \hspace{2mm}}_{\textrm{R-R}} \big\rbrace \ .
\notag
\ee
These physical scalar degrees of freedom have been identified as $\textrm{SL}(6) \times \mathbb{R}^{+}_{S} \times \mathbb{R}^{+}_{T}$ states inside the decomposition of the \textbf{133} and the results
are collected in Table~\ref{table:IIB_Gauge}.
\\

\begin{table}[t!]
\begin{center}
\scalebox{0.80}[0.85]{
\begin{tabular}{|c||c|c|c|c|}
\hline
B/F & $\sigma_{\textrm{O}3}$ & $(-1)^{F_{L}}\,\Omega_{p}$ & IIB flux & $\textrm{SL}(6) \times \mathbb{R}^{+}_{S} \times \mathbb{R}^{+}_{T}$ \\[1mm]
\hline\hline
\multirow{3}{*}{B} & $-$ & $-$ & $H_{mnp}$ & $\textbf{20}_{(-\frac{1}{2};+\frac{3}{2})}$ \\[2mm]
\cline{2-5} & $-$ & $-$ & $F_{mnp}$ & $\textbf{20}_{(+\frac{1}{2};+\frac{3}{2})}$ \\[2mm]
\hline\hline
\multirow{5}{*}{F} & $-$ & $+$ & $\partial_{m}\phi \equiv H_{m}$ & $\textbf{6'}_{(0;+1)}$ \\[2mm]
\cline{2-5} & $-$ & $+$ & ${\omega_{mn}}^{p}$ & $\textbf{84'}_{(0;+1)}$ \\[2mm]
\cline{2-5} & $-$ & $+$ & $F_{m}$ & $\textbf{6'}_{(+1;+1)}$ \\[2mm]
\cline{2-5} & $-$ & $+$ & $F_{mnpqr}$ & $\textbf{6}_{(0;+2)}$ \\[2mm]
\hline
\end{tabular}
}
\end{center}
\caption{{\it Geometric type IIB fluxes identified as states inside the decomposition of the $\textbf{912}$ of $\textrm{E}_{7(7)}$. The $ST$ weights are in perfect agreement with those ones predicted
from dimensional reduction, as shown in Appendix~\ref{App:Dim_Red}.}}
\label{table:IIB_Fluxes}
\end{table}

\noindent \textbf{The \textbf{912} representation :} This last representation of the U-duality group organises the background fluxes (generalised field strengths) threading the internal space.
These fluxes relate to the so-called embedding tensor $\,X_\mathbb{MNP}\,$ of maximal supergravity as follows \cite{Aldazabal:2010ef}
\beq
\label{partial_phi=X+...}
\partial_{\mathbb{M}} \, \phi_{\mathpzc{A}} = X_\mathbb{MNP} \, \oplus \, ... \ ,
\eeq
where the dots stand for the \textbf{56} and \textbf{6480} irep's in the product $\,\textbf{56} \times \textbf{133} = \textbf{912} + \textbf{56} + \textbf{6480}\,$, which are forbidden by
$\,\mathcal{N}=8\,$ supersymmetry \cite{deWit:2007mt}. This can be summarised as follows: \textit{the embedding tensor corresponds to the $\,\textrm{E}_{7(7)}$-variation of all the scalar fields in the 4D theory provided maximal supersymmetry is preserved}. In particular, the type IIB geometric fluxes we are considering in this work are interpreted as $\,\textrm{SL}(6)$-variations of physical fields. The different $ST$ scaling of the fluxes can be computed by dimensional reduction of the corresponding ten-dimensional Lagrangian \eqref{IIB_action} given in Appendix~\ref{App:Dim_Red}. This allows one to unambiguously identify the various IIB fluxes as states in the decomposition of the \textbf{912}. The results of this procedure are collected and shown in Table~\ref{table:IIB_Fluxes}. 

\begin{table}[t!]
\renewcommand{\arraystretch}{1.25}
\begin{center}
\scalebox{0.80}[0.85]{
\begin{tabular}{| c | c | c |}
\hline
$\textrm{SO}(6,6)$ & type IIB fluxes & isotropic couplings\\
\hline
\hline
$ -{f_{+}}^{abc} $ & $F_{ijk}$ & $ a_0 $\\
\hline
${f_{+}}^{abk}$ & $F_{ijc}$ & $ a_1 $\\
\hline
$ -{f_{+}}^{ajk}$ & $F_{ibc}$ & $ a_2 $\\
\hline
${f_{+}}^{ijk}$ & $F_{abc}$ & $ a_3 $\\
\hline
$ -{f_{-}}^{abc} $ & $ {H}_{ijk} $ & $ - b_0$\\
\hline
${f_{-}}^{abk}$ & ${H}_{ijc}$ & $- b_1 $\\
\hline
$-{f_{-}}^{ajk}$ & $ H_{a b k} $ & $ -b_2 $\\
\hline
${f_{-}}^{ijk}$ & $ H_{a b c} $ & $-b_3 $\\
\hline
\end{tabular}
\hspace{10mm}
\begin{tabular}{| c | c | c |}
\hline
$\textrm{SO}(6,6)$ & type IIB fluxes & isotropic couplings\\
\hline
\hline
$\Xi_{++\underline{a}}$ & $F_{a}$ & $ - $\\
\hline
$\Xi_{++\underline{i}}$ & $F_{i}$ & $ - $\\
\hline
\hline
${F_{a \, (0)}}$ & $ H_{a} $ & $ - $\\
\hline
${F_{i \, (0)}}$ & $ H_{i} $ & $ - $\\
\hline
$ {F^{i}}_{[\underline{bc}]}$ & ${{\omega}_{bc}}^{i}$ & $ g_{0} $\\
\hline
${F^{i}}_{[\underline{jc}]}$ & ${{\omega}_{jc}}^{i}$ & $ g_{1} $\\
\hline
${F^{a}}_{[\underline{bc}]}$ & $ {{\omega}_{bc}}^{a} $ & $ \tilde{g}_{1} $\\
\hline
${F^{a}}_{[\underline{bk}]}$ & $ {{\omega}_{bk}}^{a} $ & $ g_{2} $\\
\hline
${F^{i}}_{[\underline{j k}]} $ & ${{\omega}_{jk}}^{i}$ & $ \tilde{g}_{2} $\\
\hline
${F^{a}}_{[\underline{jk}]}$ & ${{\omega}_{jk}}^{a}$ & $ g_{3} $\\
\hline
${F_{i}}_{[\underline{jkab}]}$ & $ F_{ijk ab} $ & $ - $\\
\hline
${F_{i}}_{[\underline{jabc}]}$ & $ F_{ij abc} $ & $ - $\\
\hline
\end{tabular}
}
\end{center}
\caption{{\it Left: Mapping between orientifold-allowed geometric type IIB fluxes and bosonic embedding tensor irrep's. We have made the index splitting $\,M=(a,i,\bar{a},\bar{i})\,$ for $\textrm{SO}(6,6)$ light-cone coordinates and identified $\bar{a}$ with an upper $a$ and similarly for $\bar{i}$. \,\, Right: Mapping between orientifold-forbidden geometric type IIB fluxes and fermionic embedding tensor irrep's.
We have made the index splitting $\,\underline{m}=(\underline{a},\underline{i})\,$ for $\textrm{SL}(6)$ after using the spinor/polyform mapping described in Appendix~\ref{App:polyforms}.}}
\label{table:IIB_ET/Fluxes}
\end{table}

Alternatively to the dimensional reduction prescription, one can derive the same results by following a group theoretical approach. This entails combining derivatives and fields (see Tables~\ref{table:IIB_Der} and \ref{table:IIB_Gauge}) such that there is a complete matching of charges between the l.h.s and r.h.s of \eqref{partial_phi=X+...}. In order to obtain a precise dictionary between fluxes and embedding tensor components, we need a further breaking $\,\textrm{SO($6,6$)} \rightarrow  \textrm{SL($6$)}_{m} \rightarrow  \textrm{SL($3$)}_{a} \times\textrm{SL($3$)}_{i}\,$. This amounts to decompose the bosonic $\textrm{SO}(6,6)$ fundamental index $M$ in \textit{light-cone} coordinates as\footnote{In the rest of the paper, the index $i$ will denote an SL($3$) index in order to import results from refs~\cite{Dibitetto:2011gm,Dibitetto:2012ia} concerning fluxes. We hope not to create confusion with the SU($4$) index $i$ previously used in Section~\ref{A1A2_components}.}
\be
\label{index_splitting}
M \quad \rightarrow \quad  m \, \oplus \, \bar{m}  \quad \rightarrow \quad  a \, \oplus \, i  \, \oplus \, \bar{a}  \, \oplus \, \bar{i}  \ ,
\ee
with $\,a=\,1,3,5\,$ and  $\, i \,=\,2,4,6\,$. By using (\ref{index_splitting}) we can obtain the explicit mapping between orientifold-allowed geometric type IIB fluxes and components of $\,f_{\a MNP}\,$ and $\,\xi_{\a M}\,$ entering \eqref{Xbosonic}.  This correspondence was first found in ref.~\cite{Dibitetto:2011gm} and summarised here in Table~\ref{table:IIB_ET/Fluxes} (left). Notice that the $\xi_{\alpha M}$ piece is not activated in a geometric type IIB setup. Secondly, using the decomposition of spinorial $\textrm{SO}(6,6)$ representations given in Appendix~\ref{App:polyforms} through the mapping polyforms/spinors and further breaking the $\textrm{SL}(6)$ index $\,m \,\rightarrow\, a \,\oplus\, i\,$, one can write all those geometric type IIB fluxes which would be projected out by the orientifold projection as components of the embedding tensor pieces $\,F_{M\dm}\,$ and $\,\Xi_{\a\b\m}\,$ appearing in \eqref{Xfermionic}. This dictionary is shown in Table~\ref{table:IIB_ET/Fluxes} (right), which can be seen as the spinorial completion.

\subsection{Unorientifolding type IIA with O6-planes}
\label{sec:GroupI_embedding_IIA}

\begin{table}[t!]
\renewcommand{\arraystretch}{1.50}
\begin{center}
\scalebox{0.80}[0.85]{
\begin{tabular}{|ccl|}
\hline
$\textrm{SL}(6)$ & $\supset$ & $\textrm{SL}(3)_{a} \times \textrm{SL}(3)_{i} \times \mathbb{R}^{+}_{U}$ \\[2mm]
\hline
\hline
$\textbf{6}$ & $\rightarrow$ & $\textbf{(3,1)}_{(+\frac{1}{2})} + \textbf{(1,3)}_{(-\frac{1}{2})}$ \\[2mm]
$\textbf{15}$ & $\rightarrow$ & $\textbf{(3',1)}_{(+1)} + \textbf{(1,3')}_{(-1)} + \textbf{(3,3)}_{(0)}$ \\[2mm]
$\textbf{20}$ & $\rightarrow$ & $\textbf{(1,1)}_{(+\frac{3}{2})} + \textbf{(3',3)}_{(+\frac{1}{2})} + \textbf{(3,3')}_{(-\frac{1}{2})} + \textbf{(1,1)}_{(-\frac{3}{2})}$ \\[2mm]
$\textbf{35}$ & $\rightarrow$ & $\textbf{(1,1)}_{(0)} + \textbf{(8,1)}_{(0)} + \textbf{(1,8)}_{(0)} + \textbf{(3,3')}_{(+1)} + \textbf{(3',3)}_{(-1)}$ \\[2mm]
$\textbf{70}$ & $\rightarrow$ & $\textbf{(8,1)}_{(+\frac{3}{2})} + \textbf{(1,8)}_{(-\frac{3}{2})} + \textbf{(3,3')}_{(-\frac{1}{2})} + \textbf{(3',3)}_{(+\frac{1}{2})} + \textbf{(3,6)}_{(-\frac{1}{2})} + \textbf{(6,3)}_{(+\frac{1}{2})}$ \\[2mm]
$\textbf{84}$ & $\rightarrow$ & $\textbf{(3,1)}_{(+\frac{1}{2})} + \textbf{(1,3)}_{(-\frac{1}{2})} + \textbf{(6',1)}_{(+\frac{1}{2})} + \textbf{(1,6')}_{(-\frac{1}{2})} + \textbf{(3',3')}_{(+\frac{3}{2})} + \textbf{(3',3')}_{(-\frac{3}{2})} + \textbf{(3,8)}_{(+\frac{1}{2})} + \textbf{(8,3)}_{(-\frac{1}{2})}$ \\[2mm]
\hline
\end{tabular}
}
\end{center}
\caption{{\it Branching of $\textrm{SL}(6)$ representations according to its $\textrm{SL}(3)_{a} \times \textrm{SL}(3)_{i} \times \mathbb{R}^{+}_{U}$ subgroup. Primed irrep's have equivalent decompositions upon $\textbf{n} \leftrightarrow \textbf{n}'$ replacement and $\mathbb{R}^{+}_{U}$ sign-flip.}}
\label{table:SL6_branching}
\end{table}

As opposed to the case of type IIB with O$3$-planes, this class of type IIA backgrounds has sources which partially fill the internal space. Specifically the O$6$-planes which would break supersymmetry down to $\cN=4$ in four dimensions are placed as follows
\beq
\begin{array}{lcccc}
\textrm{O}6 \, : & & & \underbrace{\times \, \vert \, \times \, \times \, \times}_{D = 4} \, \, \times \, - \, \times \, - \, \times \, - &
\end{array}
\notag
\eeq
wrapping the internal $a=1,3,5$ directions. Unorientifolding this theory again means to place the different fluxes and fields inside bosonic or spinorial irrep's of $\textrm{SO}(6,6)$ according to whether they are allowed ($\mathbb{Z}_{2}$-even) or forbidden ($\mathbb{Z}_{2}$-odd) by the $(-1)^{F_{L}}\,\Omega_{p}\,\sigma_{\textrm{O}6}$ orientifold action. The O$6$-plane involution now reads
\beq
\sigma_{\textrm{O}6} \,\,: \,\, (\, y^{1} \, , \, y^{2} \, , \, y^{3} \, , \, y^{4} \, , \, y^{5} \, , \,y^{6} \, ) 
\,\, \rightarrow \,\, 
(\, y^{1} \, , \, -y^{2} \, , \, y^{3} \, , \, -y^{4} \, , \, y^{5} \, , \, -y^{6} \, ) \ .
\eeq

Since the $\sigma_{\textrm{O}6}$ orientifold involution breaks the $\textrm{SL}(6)$ covariance into an $\textrm{SL}(3)_{a} \times \textrm{SL}(3)_{i}\,$ one, we will need to further break the irrep's obtained in Table~\ref{table:typeII_embedding} in order to distinguish between odd and even states. Moreover, for a completely unambiguous identification, we will need the extra $\mathbb{R}^{+}_{U}$ weights treating differently $y^{a=1,3,5}$ and $y^{i=2,4,6}$, in addition to the two $\mathbb{R}^{+}$'s sitting inside $\textrm{SL}(2)_{S} \times \mathbb{R}^{+}_{T}$ which we already used in the type IIB case. The procedure followed here is, in analogy with the previous section, branching the vectors (\textbf{56}), scalars (\textbf{133}) and embedding tensor (\textbf{912}) of maximal supergravity as described in Table~\ref{table:typeII_embedding} and, subsequently further branching the results according to
\beq
\begin{array}{ccc}
\label{IIA_further_splitting}
\textrm{SL}(2)_{S} \times \textrm{SL}(6) \times \mathbb{R}^{+}_{T} & \supset & \textrm{SL}(3)_{a} \times \textrm{SL}(3)_{i} \times \mathbb{R}^{+}_{S} \times \mathbb{R}^{+}_{T} \times \mathbb{R}^{+}_{U}\ .
\end{array}
\eeq
The relevant decompositions are given in Table~\ref{table:SL6_branching}. It is worth mentioning that adopting the embedding of $\textrm{SL}(6)$ inside $\textrm{SO}(6,6)$ given in Table~\ref{table:typeII_embedding} for both type IIA and type IIB (hence named there ``type II'' embedding), is not in constrast with what found in ref.~\cite{deWit:2003hq}, where it is observed that in type IIA a different embedding is needed. This is due to the fact that essentially (up to identifications), there exists a unique decomposition once $\textrm{SL}(6)$ is further broken into $\textrm{SL}(3)_{a} \times \textrm{SL}(3)_{i}\,$. However, unlike in type IIB, the identification of the physical derivatives in the type IIA case becomes more subtle as it does not straightforwardly follow from combining the results in Tables~\ref{table:typeII_embedding} and \ref{table:SL6_branching}, as we will see next.
\\ 

\begin{table}[t!]
\begin{center}
\scalebox{0.80}[0.85]{
\begin{tabular}{|c||c|c|c|c|}
\hline
B/F & $\sigma_{\textrm{O}6}$ & $(-1)^{F_{L}}\,\Omega_{p}$ & operator & $\textrm{SL}(3)_{a} \times \textrm{SL}(3)_{i} \times \mathbb{R}^{+}_{S} \times \mathbb{R}^{+}_{T} \times \mathbb{R}^{+}_{U}$ \\[1mm]
\hline\hline
B & $+$ & $+$ & $\partial_{a}$ & $\textbf{(3,1)}_{(+\frac{1}{2};+\frac{1}{2};+\frac{1}{2})}$ \\[2mm]
\hline\hline
F & $-$ & $+$ & $\partial_{i}$ & $\textbf{(1,3')}_{(0;+1;+\frac{1}{2})}$ \\[2mm]
\hline
\end{tabular}
}
\end{center}
\caption{{\it The physical internal derivatives in type IIA compactifications. The orientifold action $(-1)^{F_{L}}\,\Omega_{p}\,\sigma_{\textrm{O}6}$ is again the combination of fermionic number, worldsheet parity and orientifold 
involution. It determines that $\partial_{a}$ is allowed by the presence of O6-planes whereas $\partial_{i}$ is not. As a consequence, they sit inside bosonic (B) and fermionic (F) irrep's of $\textrm{SO}(6,6)$, respectively.}}
\label{table:IIA_Der}
\end{table}

\noindent \textbf{The \textbf{56} representation :} The physical derivatives $\partial_{a}$ and $\partial_{i}$ are identified with the states inside the \textbf{56} displayed in Table~\ref{table:IIA_Der}. Notice that the tree physical variations $\partial_{i}$ are in common with the IIB case. In contrast, the physical variations $\partial_{a}$ have been brought from fermionic to bosonic w.r.t. the IIB case. This is consistent with the three \mbox{T-dualitites} along the $y^{1}$, $y^{3}$ and $y^{5}$ directions required to connect the IIB and the IIA duality frames.
\\

\begin{table}[t!]
\begin{center}
\scalebox{0.80}[0.85]{
\begin{tabular}{|c||c|c|c|c|}
\hline
B/F & $\sigma_{\textrm{O}6}$ & $(-1)^{F_{L}}\,\Omega_{p}$ & IIA field & $\textrm{SL}(3)_{a} \times \textrm{SL}(3)_{i} \times \mathbb{R}^{+}_{S} \times \mathbb{R}^{+}_{T} \times \mathbb{R}^{+}_{U}$ \\[1mm]
\hline\hline
\multirow{10}{*}{B} & $+$ & $+$ & $\phi$ & $\textbf{(1,1)}_{(0;0;0)}$ \\[2mm]
\cline{2-5} & $+$ & $+$ & ${e_{a}}^{b}$, ${e_{i}}^{j}$ & $\left(\textbf{(8,1)}+\textbf{(1,8)}\right)_{(0;0;0)}$ \\[2mm]
\cline{2-5} & $+$ & $+$ & ${e_{a}}^{a}$, ${e_{i}}^{i}$ & $\left(\textbf{(1,1)}+\textbf{(1,1)}\right)_{(0;0;0)}$ \\[2mm]
\cline{2-5} & $-$ & $-$ & $B_{ai}$ & $\textbf{(3,3')}_{(0;0;+1)}$ \\[2mm]
\cline{2-5} & $-$ & $-$ & $C_{i}$ & $\textbf{(1,3')}_{(0;+1;-1)}$ \\[2mm]
\cline{2-5} & $+$ & $+$ & $C_{abc}$ & $\textbf{(1,1)}_{(+1;0;0)}$ \\[2mm]
\cline{2-5} & $+$ & $+$ & $C_{ajk}$ & $\textbf{(3,3)}_{(0;+1;0)}$ \\[2mm]
\cline{2-5} & $-$ & $-$ & $C_{abijk}$ & $\textbf{(3',1)}_{(0;+1;+1)}$ \\[2mm]
\hline\hline
\multirow{10}{*}{F} & $-$ & $+$ & ${e_{a}}^{i}$, ${e_{i}}^{a}$ & $\textbf{(3,3)}_{(+\frac{1}{2};-\frac{1}{2};0)} + \textbf{(3',3')}_{(-\frac{1}{2};+\frac{1}{2};0)}$ \\[2mm]
\cline{2-5} & $+$ & $-$ & $B_{ab}$ & $\textbf{(3',1)}_{(+\frac{1}{2};-\frac{1}{2};+2)}$ \\[2mm]
\cline{2-5} & $+$ & $-$ & $B_{ij}$ & $\textbf{(1,3)}_{(-\frac{1}{2};+\frac{1}{2};+2)}$ \\[2mm]
\cline{2-5} & $-$ & $+$ & $B_{abcijk}$ & $\textbf{(1,1)}_{(+\frac{1}{2};+\frac{3}{2};0)}$ \\[2mm]
\cline{2-5} & $+$ & $-$ & $C_{a}$ & $\textbf{(3,1)}_{(+\frac{1}{2};+\frac{1}{2};-1)}$ \\[2mm]
\cline{2-5} & $-$ & $+$ & $C_{abk}$ & $\textbf{(3',3')}_{(+\frac{1}{2};+\frac{1}{2};0)}$ \\[2mm]
\cline{2-5} & $-$ & $+$ & $C_{ijk}$ & $\textbf{(1,1)}_{(-\frac{1}{2};+\frac{3}{2};0)}$ \\[2mm]
\cline{2-5} & $+$ & $-$ & $C_{abcij}$ & $\textbf{(1,3)}_{(+\frac{1}{2};+\frac{1}{2};1)}$ \\[2mm]
\hline
\end{tabular}
}
\end{center}
\caption{{\it The physical scalars from type IIA compactifications mapped into states in the $\textbf{133}$ of $\textrm{E}_{7(7)}$. Note that it is the combination of fermionic number, worldsheet 
parity and orientifold involution what determines which states are bosonic (B) and fermionic (F). It is worth mentioning that, in order to get the correct number of physical degrees of freedom (\emph{i.e.}
$70 = 38_{\textrm{B}} + 32_{\textrm{F}}$), one needs to subtract the compact directions inside the vielbein.}}
\label{table:IIA_Gauge}
\end{table}

\noindent \textbf{The \textbf{133} representation :} We will again identify the physical scalars (which carry $70$ degrees of freedom in total) with the pure scalars coming from the democratic 10D fields in type IIA supergravity \cite{Bergshoeff:2001pv} having all legs threading the internal space. These $70$ scalars split up into $38$ orientifold-allowed ones arising from
\be
\label{IIA_B_scalars}
\big\lbrace \underbrace{\hspace{2mm} \phi \hspace{3.5mm} , \hspace{4mm} {e_{a}}^{b} \hspace{3.5mm} , \hspace{3.5mm} {e_{i}}^{j} \hspace{3.5mm} , \hspace{3.5mm} {e_{a}}^{a} \hspace{3.5mm} , \hspace{3.5mm}
{e_{i}}^{i} \hspace{3.5mm} , \hspace{3.5mm} B_{ai} \hspace{3.5mm}\hspace{2mm}}_{\textrm{NS-NS}} \hspace{3.5mm} ,
\hspace{3.5mm} \underbrace{\hspace{2mm} C_{i} \hspace{3.5mm} , \hspace{3.5mm} C_{abc} \hspace{3.5mm} , \hspace{3.5mm} C_{ajk} \hspace{3.5mm} , \hspace{3.5mm} C_{abijk}
\hspace{2mm}}_{\textrm{R-R}} \big\rbrace \ ,
\notag
\ee
where the correct counting is reproduced upon subtracting the $6$ compact $\textrm{SO}(3) \times \textrm{SO}(3)$ directions inside the vielbeins, and $32$ orientifold-forbidden ones coming from
\be
\label{IIA_F_scalars}
\big\lbrace \underbrace{\hspace{2mm} {e_{a}}^{i} \hspace{3.5mm} , \hspace{3.5mm} {e_{i}}^{a} \hspace{3.5mm} , \hspace{3.5mm} B_{ab} \hspace{3.5mm} , \hspace{3.5mm} B_{ij} \hspace{3.5mm} ,
\hspace{3.5mm} B_{abcijk} \hspace{2mm}}_{\textrm{NS-NS}} \hspace{3.5mm} ,
\hspace{3.5mm} \underbrace{\hspace{2mm} C_{a} \hspace{3.5mm} , \hspace{3.5mm} C_{abk} \hspace{3.5mm} , \hspace{3.5mm} C_{ijk} \hspace{3.5mm} , \hspace{3.5mm} C_{abcij} \hspace{2mm}}_{\textrm{R-R}}
\big\rbrace \ ,
\notag
\ee
where, now one should subtract $9$ compact vielbein directions to get the correct counting. The above scalars can be traced back to the corresponding states in the decomposition of the \textbf{133} in Table~\ref{table:typeII_embedding} by using the branching (\ref{IIA_further_splitting}) and the results collected in Table~\ref{table:IIA_Gauge}.
\\[4mm]

\begin{table}[t!]
\begin{center}
\scalebox{0.80}[0.85]{
\begin{tabular}{|c||c|c|c|c|}
\hline
B/F & $\sigma_{\textrm{O}6}$ & $(-1)^{F_{L}}\,\Omega_{p}$ & IIA flux & $\textrm{SL}(3)_{a} \times \textrm{SL}(3)_{i} \times \mathbb{R}^{+}_{S} \times \mathbb{R}^{+}_{T} \times \mathbb{R}^{+}_{U}$ \\[1mm]
\hline\hline
\multirow{12}{*}{B} & $+$ & $+$ & $\partial_{a}\phi \equiv H_{a}$ & $\textbf{(3,1)}_{(+\frac{1}{2};+\frac{1}{2};+\frac{1}{2})}$ \\[2mm]
\cline{2-5} & $+$ & $+$ & ${\omega_{ij}}^{c}$ & $\textbf{(3',3)}_{(-\frac{1}{2};+\frac{3}{2};+\frac{1}{2})}$ \\[2mm]
\cline{2-5} & $+$ & $+$ & ${\omega_{aj}}^{k}$, ${\omega_{ab}}^{c}$ & $\left(\textbf{(3,8)}+\textbf{(6',1)}\right)_{(+\frac{1}{2};+\frac{1}{2};+\frac{1}{2})}$ \\[2mm]
\cline{2-5} & $-$ & $-$ & $H_{ijk}$ & $\textbf{(1,1)}_{(-\frac{1}{2};+\frac{3}{2};+\frac{3}{2})}$ \\[2mm]
\cline{2-5} & $-$ & $-$ & $H_{abk}$ & $\textbf{(3',3')}_{(+\frac{1}{2};+\frac{1}{2};+\frac{3}{2})}$ \\[2mm]
\cline{2-5} & $-$ & $-$ & $F_{aibjck}$ & $\textbf{(1,1)}_{(+\frac{1}{2};+\frac{3}{2};+\frac{3}{2})}$ \\[2mm]
\cline{2-5} & $+$ & $+$ & $F_{aibj}$ & $\textbf{(3',3)}_{(+\frac{1}{2};+\frac{3}{2};+\frac{1}{2})}$ \\[2mm]
\cline{2-5} & $-$ & $-$ & $F_{ai}$ & $\textbf{(3,3')}_{(+\frac{1}{2};+\frac{3}{2};-\frac{1}{2})}$ \\[2mm]
\cline{2-5} & $+$ & $+$ & $F_{0}$ & $\textbf{(1,1)}_{(+\frac{1}{2};+\frac{3}{2};-\frac{3}{2})}$ \\[2mm]
\hline\hline
\multirow{12}{*}{F} & $-$ & $+$ & $\partial_{i}\phi \equiv H_{i}$ & $\textbf{(1,3')}_{(0;+1;+\frac{1}{2})}$ \\[2mm]
\cline{2-5} & $-$ & $+$ & ${\omega_{ab}}^{k}$ & $\textbf{(3',3)}_{(+1;0;+\frac{1}{2})}$ \\[2mm]
\cline{2-5} & $-$ & $+$ & ${\omega_{ij}}^{k}$, ${\omega_{ib}}^{c}$ & $\left(\textbf{(1,6)}+\textbf{(8,3')}\right)_{(0;+1;+\frac{1}{2})}$ \\[2mm]
\cline{2-5} & $+$ & $-$ & $H_{abc}$ & $\textbf{(1,1)}_{(+1;0;+\frac{3}{2})}$ \\[2mm]
\cline{2-5} & $+$ & $-$ & $H_{ajk}$ & $\textbf{(3,3)}_{(0;+1;+\frac{3}{2})}$ \\[2mm]
\cline{2-5} & $-$ & $+$ & $F_{aijk}$ & $\textbf{(3,1)}_{(0;+2;+\frac{1}{2})}$ \\[2mm]
\cline{2-5} & $-$ & $+$ & $F_{abci}$ & $\textbf{(1,3')}_{(+1;+1;+\frac{1}{2})}$ \\[2mm]
\cline{2-5} & $+$ & $-$ & $F_{ab}$ & $\textbf{(3',1)}_{(+1;+1;-\frac{1}{2})}$ \\[2mm]
\cline{2-5} & $+$ & $-$ & $F_{ij}$ & $\textbf{(1,3)}_{(0;+2;-\frac{1}{2})}$ \\[2mm]
\hline
\end{tabular}
}
\end{center}
\caption{{\it Geometric type IIA fluxes identified as states inside the decomposition of the $\textbf{912}$ of $\textrm{E}_{7(7)}$. The $STU$ weights are in perfect agreement with those ones predicted
from dimensional reduction, as shown in Appendix~\ref{App:Dim_Red}.}}
\label{table:IIA_Fluxes}
\end{table}

\noindent \textbf{The \textbf{912} representation :} Let us conclude this section by exploring the different deformations of maximal supergravity in its type IIA incarnation. The $STU$ weights of all the geometric type IIA fluxes can be obtained
by dimensional reduction of the corresponding terms in the ten-dimensional \textit{massive} IIA Lagrangian, as explained in Appendix~\ref{App:Dim_Red}, and then unambiguously identified inside the
$\textrm{SL}(3)_{a} \times \textrm{SL}(3)_{i} \times \mathbb{R}^{+}_{S} \times \mathbb{R}^{+}_{T} \times \mathbb{R}^{+}_{U}$ decomposition of the $\textbf{912}$. This prescription works in complete analogy to the type IIB case and the results are summarised in Table~\ref{table:IIA_Fluxes}.

\begin{table}[t!]
\renewcommand{\arraystretch}{1.25}
\begin{center}
\scalebox{0.80}[0.85]{
\begin{tabular}{| c | c | c |}
\hline
$\textrm{SO}(6,6)$ & type IIA fluxes & isotropic couplings\\
\hline
\hline
$ -{f_{+}}^{abc} $ & $F_{aibjck}$ & $ a_0 $\\
\hline
${f_{+}}^{abk}$ & $F_{aibj}$ & $ a_1 $\\
\hline
$ -{f_{+}}^{ajk}$ & $F_{ai}$ & $ a_2 $\\
\hline
${f_{+}}^{ijk}$ & $F_{0}$ & $ a_3 $\\
\hline
$ -{f_{-}}^{abc} $ & $ {H}_{ijk} $ & $ - b_0$\\
\hline
${f_{-}}^{abk}$ & ${{\omega}_{ij}}^{c}$ & $ - b_1 $\\
\hline
${{f_{+}}^{ab}}_{k}$ & $ H_{a b k} $ & $ c_0 $\\
\hline
${{f_{+}}^{aj}}_{k}$ & $ {\omega_{a k}}^{j}$ & $c_1 $\\
\hline
${{f_{+}}^{ab}}_{c}$ & $ {\omega_{a b}}^{c} $ & $\tilde {c}_1 $\\
\hline
\hline
$ {\xi_{+}}^{a} $ & $H_{a}$ & $ - $\\
\hline
\end{tabular}
\hspace{10mm}
\begin{tabular}{| c | c | c |}
\hline
$\textrm{SO}(6,6)$ & type IIA fluxes & isotropic couplings\\
\hline
\hline
$\Xi_{++\underline{c}}$ & $\frac{1}{2} \, \epsilon^{abc} \, F_{ab}$ & $ - $\\
\hline
$\Xi_{++\underline{i}}$ & $ F_{abci} $ & $ - $\\
\hline
$\Xi_{+-\underline{i}}$ & $ H_{i}$ & $ - $\\
\hline
${\Xi_{++}}^{[\underline{abc}]}$ & $ H_{abc} $ & $-h^{+}_{0}$\\
\hline
$ {\Xi_{++}}^{[\underline{abk}]} $ & ${{\omega}_{ab}}^{k}$ & $ h^{+}_{1} $\\
\hline
\hline
${F^{d}}_{[\underline{aibjck}]} $ & $F_{dijk}$ & $ - $\\
\hline
${F^{l}}_{[\underline{aibjck}]}$ & $\frac{1}{2} \, F_{ij} \, \epsilon^{ijl}$ & $ - $\\
\hline
${F^{b}}_{[\underline{ci}]}$ & ${{\omega}_{bi}}^{c}$ & $ g_{2} $\\
\hline
$ {F^{k}}_{[\underline{ij}]}$ & ${{\omega}_{ij}}^{k}$ & $ \tilde{g}_{2} $\\
\hline
${F^{a}}_{[\underline{jk}]}$ & $ {H}_{ajk} $ & $ g_{3} $\\
\hline
\end{tabular}

}
\end{center}
\caption{{\it Left: Mapping between orientifold-allowed geometric type IIA fluxes and bosonic embedding tensor irrep's.
We have made the index splitting $\,M=(a,i,\bar{a},\bar{i})\,$ for $\textrm{SO}(6,6)$ light-cone coordinates and identified $\bar{a}$ with an upper $a$ and similarly for $\bar{i}$.   \,\, Right: Mapping between orientifold-forbidden geometric type IIA fluxes and fermionic embedding tensor irrep's. We have made the index splitting $\,\underline{m}=(\underline{a},\underline{i})\,$ for $\textrm{SL}(6)$ after using the spinor/polyform mapping described in Appendix~\ref{App:polyforms}.}}
\label{table:IIA_ET/Fluxes}
\end{table}

However, there is a fundamental obstruction to derive the same results by following the group theoretical approach of matching charges between the l.h.s and r.h.s of (3.5) using only geometric ingredients: in the type IIA case, the Romans' mass parameter $F_{0}$ cannot be obtained as the $\textrm{SL}(6)$-variation of a physical field. 
This mismatch is simply due to the fact that $F_{0}$ is already a consistent deformation of the original theory in 10D and it does not originate from any internal dependence of the fields upon dimensional
reduction. This deformation parameter corresponds to the state $\textbf{(1,1)}_{(+\frac{1}{2};+\frac{3}{2};-\frac{3}{2})}$ in Table~\ref{table:IIA_Fluxes}. Then, by inspection of Tables~{\ref{table:IIA_Der}} and \ref{table:IIA_Gauge}, one gets quickly convinced that this state cannot be generated in a geometric way. Nevertheless, if one insists on the embedding tensor still being the E$_{7(7)}$-variation of all the scalar fields in the 4D theory provided maximal supersymmetry is preserved, then one can look for the candidate to be the Romans' mass according to group theory. The answer is given by
\beq
\label{Roman_mass}
\begin{array}{cccccccc}
F_{0} &\equiv & \textbf{(1,1)}_{(+\frac{1}{2};+\frac{3}{2};-\frac{3}{2})} &=& 
\textbf{(1,3)}_{(+\frac{1}{2};+\frac{1}{2};-\frac{1}{2})} &\times & \textbf{(1,3')}_{(0;+1;-1)} & \\[2mm]
&  & & + & \textbf{(3',1)}_{(0;+1;-\frac{1}{2})} &\times & \textbf{(3,1)}_{(+\frac{1}{2};+\frac{1}{2};-1)} & \\[2mm]
&  & & + & \textbf{(1,1)}_{(0;0;-\frac{3}{2})} &\times & \textbf{(1,1)}_{(+\frac{1}{2};+\frac{3}{2};0)} & , \\[-2mm]
&  &  &          &   \underbrace{\textrm{ \phantom{a} \hspace{20mm} \phantom{a} }}_{\textrm{E}_{7(7)}/\textrm{SL}(6)\textrm{-variations}}         &             &    \underbrace{\textrm{ \phantom{a} \hspace{20mm} \phantom{a} }}_{\textrm{type IIA fields }}  &        
\end{array}
\eeq
providing an interpretation of the 10D Romans' deformation in the 4D context of EGG. More concretely, the parameter $F_{0}$ is associated to variations beyond the $\textrm{SL}(6)$-type\footnote{This is in
 line with ref.~\cite{Hohm:2011cp}, where massive type IIA supergravity was obtained by means of a twisted reduction of double field theory upon including some non-trivial dependence on dual coordinates violating 
the strong constraint.} of the physical fields $C_{i}\equiv \textbf{(1,3')}_{(0;+1;-1)}$, $C_{a}\equiv \textbf{(3,1)}_{(+\frac{1}{2};+\frac{1}{2};-1)}$  and $B_{abcijk}\equiv \textbf{(1,1)}_{(+\frac{1}{2};+\frac{3}{2};0)}$ (first, second and third line in (\ref{Roman_mass}), respectively). Therefore, according to the definition of geometric fluxes adopted in ref.~\cite{Aldazabal:2010ef}, \textit{i.e.} $\textrm{SL}(6)$-variations of physical fields, the Romans' mass represents a non-geometric flux in 4D (not even locally geometric) with the higher-dimensional interpretation of a deformation parameter already in 10D.

\begin{figure}[t!]
\begin{center}
\includegraphics[width=125mm]{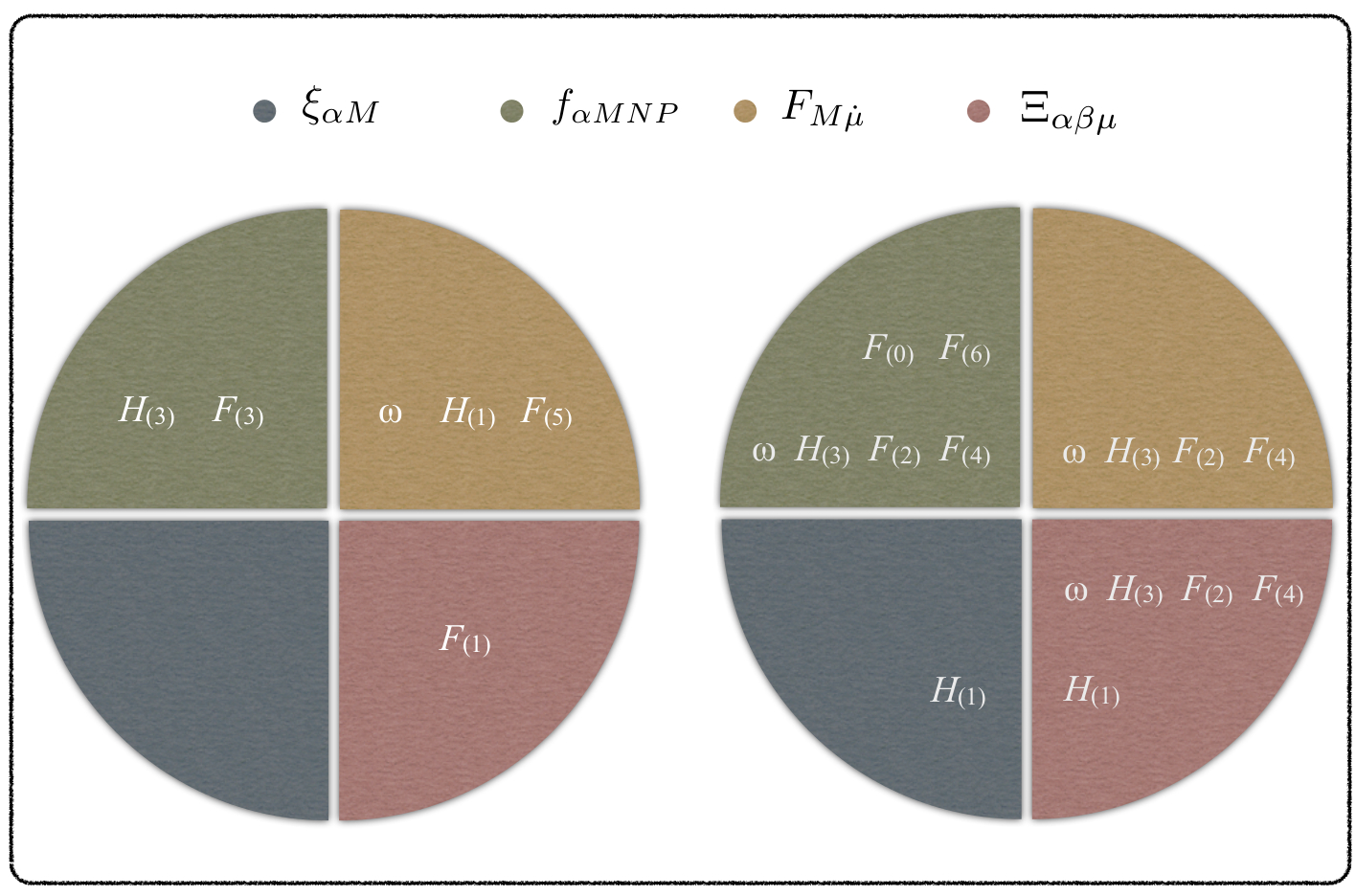}
\caption{{\it Distribution of type IIB (left) and type IIA (right) fluxes along the different embedding tensor pieces. As one can see, type IIA compactifications tend to spread all the fluxes much more
than type IIB and activate different embedding tensor irrep's, thus generating a larger variety of moduli dependences in the flux-induced scalar potential.}}
\label{Fig:ET/fluxes} 
\end{center}
\end{figure}

Finally, splitting again the $\textrm{SO}(6,6)$ index $M$ in \textit{light-cone} coordinates according to \eqref{index_splitting}, we obtain the explicit mapping between orientifold-allowed geometric type IIA fluxes -- as components of $f_{\a MNP}$ and $\xi_{\a M}$ -- summarised in Table~\ref{table:IIA_ET/Fluxes} (left). In addition, using the polyforms/spinors mapping and the SL($6$) index splitting $\,m \, \rightarrow \,a \, \oplus \,i\,$, we can determine all the geometric type IIA fluxes sitting inside the $\,F_{M\dm}\,$ and $\,\Xi_{\a\b\m}\,$ pieces which would be projected out by the orientifold action. We have summarised the results in Table~\ref{table:IIA_ET/Fluxes} (right). 
Remarkably, all the pieces of the embedding tensor are activated in a geometric type IIA setup. For the sake of clarity, we have depicted this situation in Figure~\ref{Fig:ET/fluxes}, where the difference
between type IIB and IIA is highlighted in this respect.

\section{Testing the fluxes/ET correspondence}
\label{sec:Vacua}

In the previous sections we have derived the precise correspondence between type II fluxes (both IIB and IIA), the set of embedding tensor components $\,f_{\alpha MNP}\,$, $\,\xi_{\alpha M}\,$, $\,F_{M \dot{\mu}}\,$ and $\,\Xi_{\alpha\beta \mu}$ and the fermi mass terms $\mathcal{A}^{\mathcal{IJ}}$ and ${\mathcal{A}_{\mathcal{I}}}^{\mathcal{JKL}}$. Here we will test this correspondence using a twisted $\mathbb{T}^{6}/(\mathbb{Z}_{2} \times \mathbb{Z}_{2})$ isotropic\footnote{In the supergravity language, working with this isotropic orbifold amounts to consider the SO(3)-invariant sector of maximal supergravity. This sector preserves $\mathcal{N}=2$ supersymmetry and the scalars span the coset space $\frac{\textrm{SL}(2)}{\textrm{SO}(2)} \times \frac{\textrm{G}_{2(2)}}{\textrm{SO}(4)}$, which can be viewed as a submanifold inside the full $\frac{\textrm{E}_{7(7)}}{\textrm{SU}(8)}$ scalar manifold. Restricting to the $\mathbb{Z}_{2}$ orientifold-even subsector further reduces the scalar manifold to an $\left( \frac{\textrm{SL}(2)}{\textrm{SO}(2)} \right)^{3}$ coset space and the resulting supergravity still preserves $\mathcal{N}=1$ supersymmetry \cite{Dibitetto:2012ia}.} orbifold compactification as playground\footnote{We refer the reader to ref.~\cite{Guarino:2010zz} for a detailed description of our conventions concerning the $\mathbb{Z}_{2} \times \mathbb{Z}_{2}$ orbifold geometry.} and will chart the landscape of the resulting $\mathcal{N}=8$ gauged supergravities. We will present the explicit form of the quadratic constraints in (\ref{quadratic_const}) in terms of the geometric type II fluxes in the tables and will interpret them as the vanishing of the flux-induced tadpoles for the different sources absent in our setup. As we will see, the situation is different in type IIA and IIB scenarios. In the former case, the set of sources for which a tadpole can be induced after turning on spinorial geometric fluxes is the same as in the bosonic setup. In the latter, odd fluxes induce tadpoles for more types of sources than their bosonic counterparts. Subsequently, we will go through the analysis of critical points in the two different cases. 

Before attacking that problem, though, we will make use of the symmetries of the corresponding scalar potentials
in order to simplify the analysis as much as we can. First of all, both in the type IIA and in the type IIB case, the set of geometric fluxes which we turn on happens to be a closed set under non-compact $\textrm{E}_{7(7)}$ tranformations\footnote{The non-compact transformations needed to bring the ten physical scalars in the $\textrm{SL}(2)/\textrm{SO}(2) \times \textrm{G}_{2(2)}/\textrm{SO}(4)$ scalar manifold to the origin correspond to the three Cartan's and the seven positive roots \cite{Ferrara:1989ik,deWit:1991pa}. This is analogous to the orientifolded case where the three Cartan's and the three positive roots are needed to bring to the origin the six physical scalars in $\left(\textrm{SL}(2)/\textrm{SO}(2)\right)^{3}$ \cite{Dibitetto:2011gm}.}. Hence, one can exhaustively restrict the search for critical points to the origin of moduli space $\,\phi_{\mathpzc{A}}=0\,$, where the EOM's in (\ref{scalars_eom}) take the simple form of algebraic quadratic equations in the fluxes \cite{Dibitetto:2011gm,DallAgata:2011aa}. Secondly, since the origin of moduli space is left invariant by the action of compact $\textrm{SU}(8)$ transformations, the EOM's will have an extra residual compact symmetry provided that the corresponding set of fluxes is closed under such compact duality transformations as well. We would like to stress that this will not be the case in general and such transformations will typically turn on non-geometric fluxes. Making use of a particular compact residual symmetry will be, in what follows, referred to as \emph{gauge fixing}.

\subsection{Type IIB without O3-planes}

Let us start by testing the fluxes/embedding tensor correspondence in the case of a type IIB flux compactification on a twisted $\mathbb{T}^{6}/(\mathbb{Z}_{2} \times \mathbb{Z}_{2})$ orbifold. In this case one would expect to find tadpole cancellation conditions involving O3/D3, O5/D5 and O7/D7 systems coming from the consistency condition (\ref{quadratic_const}). However, a flux-induced tadpole for the $C_{8}$ gauge potential cannot be induced unless certain non-geometric fluxes are included in the construction \cite{Aldazabal:2006up}. In this work we restric ourselves to geometric setups, so only flux-induced tadpoles of the form
\beq
\label{C4_C6_tadpole}
\int ( H_{3} \wedge F_{3} ) \wedge C_{4}
\hspace{8mm} \textrm{ and } \hspace{8mm}
\int \omega F_{3}  \wedge C_{6}
\eeq
will appear and again will potentially induce quadratic relations on the set of type IIB fluxes. 

The case of geometric isotropic type IIB compactifications consists of $14$ fluxes displayed in the right column of Table~\ref{table:IIB_ET/Fluxes}: R-R fluxes ($a_{0}$, $a_{1}$, $a_{2}$, $a_{3}$), NS-NS fluxes ($b_{0}$, $b_{1}$, $b_{2}$, $b_{3}$) and metric fluxes ($g_{0}$, $g_{1}$, $\tilde{g}_{1}$, $g_{2}$, $\tilde{g}_{2}$, $g_{3}$).

\subsubsection{Quadratic constraints and sources}

Plugging the set of geometric IIB fluxes in Table~\ref{table:IIB_ET/Fluxes} into the quadratic constraints in (\ref{quadratic_const}) produces the following set of conditions:
\begin{itemize}

\item Nilpotency ($D^2=0$) of the $D = d+\omega$ operator in the internal space: This condition can be written as $\,{\omega_{[m_{1} m_{2}}}^{p}\, {\omega_{m_{3} ] p}}^{m_{4}}=0\,$ and produces three independent relations on the fluxes
\beq
\label{IIB_nilpotency}
\begin{array}{rccc}
g_1 \, (g_1-\tilde{g}_{1}) \,+\, g_{0} \, (g_{2}-\tilde{g}_{2}) & = & 0 & , \\
g_{2} \, (g_{2}-\tilde{g}_{2}) \,+\, g_3 \, (g_1-\tilde{g}_{1}) & = & 0 & , \\
g_{1} \, g_{2} \,-\, g_{0} \, g_3 & = & 0 & ,
\end{array}
\eeq
which can be interpreted as requiring the absence of KK$5$-branes \cite{Villadoro:2007yq}.

\item Closure of $H_{3}$ under $D$: This condition can be expressed as ${\omega_{[m_{1} m_{2}}}^{p}\, H_{m_{3} m_{4}] p}=0$ and yields the following condition on the fluxes
\beq
g_{0} \, b_{0} \,-\, (2 g_{1} - \tilde{g}_{1}) \, b_{1} \,-\, (2 g_{2} - \tilde{g}_{2}) \, b_{2} \,+\, g_{3} \, b_{3} = 0 \ ,
\eeq
which is equivalent to demanding the absence of NS$5$-branes \cite{Villadoro:2007yq,Villadoro:2007tb}.

\item Tadpole cancellation condition for the $C_{4}$ gauge potential due to the topological term in (\ref{C4_C6_tadpole}). It produces a single relation associated to $\,H_{[m_{1} m_{2} m_{3}} \, F_{m_{4} m_{5} m_{6} ]} =0\,$, namely,
\beq
\label{No_D3}
b_{3} \, a_{0} \,-\, 3 \, b_{2} \, a_{1} \,+\, 3 \, b_{1} \, a_{2} \,-\, b_{0} \, a_{3} =0 \ .
\eeq

\item Tadpole cancellation condition for the $C_{6}$ gauge potential displayed in (\ref{C4_C6_tadpole}). There is a single relation coming from ${\omega_{[m_{1} m_{2}}}^{p} F_{m_{3} m_{4}] p}=0$, which reads
\beq
\label{No_D5}
g_{0} \, a_{0} \,-\, (2 g_{1} - \tilde{g}_{1}) \, a_{1} \,-\, (2 g_{2} - \tilde{g}_{2}) \, a_{2} \,+\, g_{3} \, a_{3} = 0 \ .
\eeq

\end{itemize}

The above set of consistency relations nicely generalises the bosonic results in ref.~\cite{Dibitetto:2011gm}. Notice that only the  tadpole cancellation condition for $C_{4}$ survives in a purely bosonic IIB setup where the metric flux (parity-odd under the orientifold action) is absent.

\subsubsection{The IIB landscape}

The EOM's at the origin of the moduli space can be obtained by plugging the expressions (\ref{A1even})--(\ref{A2odd}) for the fermion mass terms as a function of the embedding tensor pieces into (\ref{scalars_eom}) and then using the identification in Table~\ref{table:IIB_ET/Fluxes} between embedding tensor components and type IIB fluxes. The result is then a set of quadratic relations on the fluxes which still has to be supplemented with those in (\ref{IIB_nilpotency})--(\ref{No_D5}) coming from the consistency of the flux-induced gauging in $\mathcal{N}=8$.

We will fix the gauge by setting $b_{0}=\tilde{g}_{2}=0$. This can be carried out by first using $\textrm{SO}(2)_{U}$, w.r.t. which the whole set of fluxes in Table~\ref{table:IIB_ET/Fluxes} is manifestly invariant. Subsequently, one can still make use of $\textrm{SO}(2)_{S}$ (under which all the components of metric flux do not transform) to set $b_{0}=0$. After some algebra manipulations, it can be shown that the system of equations combining (\ref{IIB_nilpotency})--(\ref{No_D5}) + EOM's at the origin, demands a vanishing metric flux, \textit{i.e.} ($g_{0}$, $g_{1}$, $\tilde{g}_{1}$, $g_{2}$, $\tilde{g}_{2}$, $g_{3}$) $=0$.
Therefore, there are no solutions in the geometric IIB even after including fermi fluxes. 

It is worth mentioning that the only known (isotropic) solutions within geometric type IIB compactifications with
only gauge fluxes are of the GKP-type \cite{Giddings:2001yu} and crucially require the presence of O$3$-planes to cancel the flux-induced tadpole for the $C_{4}$ potential in (\ref{C4_C6_tadpole}).

\subsection{Type IIA without O6-planes}

Now we will test the fluxes/embedding tensor correspondence in the case of a type IIA flux compactification also on a twisted $\mathbb{T}^{6}/(\mathbb{Z}_{2} \times \mathbb{Z}_{2})$ orbifold. Since this orbifold is a Calabi-Yau space far from the singularities, systems of O4/D4 and O8/D8 sources are not allowed due to the absence of 1-cycles and 5-cycles \cite{Grana:2005jc}. Consequently, the quadratic constraints in (\ref{quadratic_const}) are not expected to reproduce tadpole cancellation conditions involving these types of localised sources. On the other hand, a flux-induced tadpole for the R-R field $C_{7}$ of the form
\beq
\label{C7_tadpole}
\int ( \omega \,  F_{2} + H_{3} \, F_{0} ) \wedge \, C_{7}
\eeq
will still be produced yielding algebraic constraints on the flux parameters \cite{Camara:2005dc}. More concretely, there will be four of such relations associated to the four independent 3-cycles in the $\mathbb{Z}_{2} \times \mathbb{Z}_{2}$ isotropic orbifold.

The geometric flux content in isotropic type IIA compactifications consists of the $14$ fluxes displayed in the last column of Table~\ref{table:IIA_ET/Fluxes} : R-R fluxes ($a_{0}$, $a_{1}$, $a_{2}$, $a_{3}$), NS-NS fluxes ($b_{0}$, $g_{3}$, $c_{0}$, $h^{+}_{0}$) and metric fluxes ($h^{+}_{1}$, $c_{1}$, $\tilde{c}_{1}$, $g_{2}$, $\tilde{g}_{2}$, $b_{1}$).

\subsubsection{Quadratic constraints and sources}

Proceeding in an analogous manner as in the type IIB case,  the consistency requirement in eq.~(\ref{quadratic_const}) produces the following set of conditions:
\begin{itemize}

\item Nilpotency ($D^2=0$) of the $D = d+\omega$ operator in the internal space: As before, this condition yields three independent relations on the fluxes
\beq
\begin{array}{rccc}
\label{IIA_nilpotency}
c_1 \, (c_1-\tilde{c}_{1}) \,+\, h^{+}_{1}\, (g_{2}-\tilde{g}_{2}) & = & 0 & , \\
g_{2} \, (g_{2}-\tilde{g}_{2}) \,-\, b_1 \, (c_1-\tilde{c}_{1}) & = & 0 & , \\
c_{1} \, g_{2} \,-\, h^{+}_{1} \, b_1 & = & 0 & .
\end{array}
\eeq

\item Closure of $H_{3}$ under $D$: This time, it gives rise to the flux relation
\beq
\label{IIA_closure_H3}
b_{0} \, h^{+}_{1} \,+\, g_{3} \, (2 c_{1} - \tilde{c}_{1}) \,+\, c_{0} \, (2 g_{2} - \tilde{g}_{2}) \,-\, b_{1} \, h^{+}_{0} = 0 \ .
\eeq

\item Tadpole cancellation conditions for $C_{7}$ corresponding to the different components in ${\omega_{[m_{1} m_{2}}}^{p}\, F_{m_{3}] p} + H_{m_{1} m_{2} m_{3} } \, F_{0}=0$ coming from the topological term (\ref{C7_tadpole}). These are given by
\beq
\label{NO_D6}
\begin{array}{rlrccc}
\left[ ijk \right] \, \textrm{ component} & : \hspace{10mm}& 3 \, b_{1} \, a_{2} \,-\, b_{0} \, a_{3} & = & 0 & , \\
\left[ ijc \right] \, \textrm{ component} & : & (2 g_{2} \,-\, \tilde{g}_{2}) \, a_{2} \,-\, g_{3} \, a_{3} & = & 0 & , \\
\left[ ibc \right] \, \textrm{ component} & : & (2 c_{1} - \tilde{c}_{1}) \, a_{2} \,+\, c_{0} \, a_{3} & = & 0 & , \\
\left[ abc \right] \, \textrm{ component} & : & 3 \, h^{+}_{1} \, a_{2} \,-\, h^{+}_{0} \, a_{3} & = & 0 & .
\end{array}
\eeq
Notice that the first and the third conditions are parity-even with respect to the orientifold action and thus were already present in the bosonic setup, whereas the second and the fourth conditions
vanish in a purely bosonic setup. The constraints collected in \eqref{NO_D6} imply the absence of D$6$-branes.

\end{itemize}

As for the type IIB case, the above set (\ref{IIA_nilpotency})--(\ref{NO_D6}) of quadratic constraints nicely generalises the previous bosonic results in ref.~\cite{Dibitetto:2011gm}.

\begin{table}[t!]
\renewcommand{\arraystretch}{1.80}
\begin{center}
\scalebox{0.77}[0.77]{
\begin{tabular}{ | c || c | c |c | c | c | c |c | c || c |}
\hline
\textrm{\textsc{id}} & $a_{0}$ & $a_{1}$ & $a_{2}$ & $a_{3}$ & $b_{0}$ & $b_{1}$ & $c_{0}$ & $c_{1}=\tilde{c}_{1}$ & $V_{0}$  \\[1mm]
\hline \hline
$1$ & $\dfrac{3 \,\sqrt{10}}{2}\, \lambda $ & $\dfrac{\sqrt{6}}{2} \, \lambda$ & $ - \dfrac{\sqrt{10}}{6} \, \lambda$ & $\dfrac{5\,\sqrt{6}}{6} \, \lambda$ & $-\dfrac{\sqrt{6}}{3} \, \lambda$ & $\dfrac{\sqrt{10}}{3}\,\lambda$ & $\dfrac{\sqrt{6}}{3}\,\lambda$ & $\sqrt{10} \, \lambda$  & $-\lambda^{2}$  \\[1mm]
\hline \hline
$2$ & $\dfrac{16 \, \sqrt{10}}{9} \,\lambda$ & $0$ & $0$ & $\dfrac{16 \, \sqrt{2}}{9} \, \lambda$ & $0$ & $\dfrac{16 \, \sqrt{10}}{45} \, \lambda$ & $0$ & $\dfrac{16 \, \sqrt{10}}{15} \, \lambda$  & $-\dfrac{32}{27} \, \lambda^{2}$ \\[1mm]
\hline
$3$ & $\dfrac{4\,\sqrt{10}}{5}\,\lambda$ & $-\dfrac{4\,\sqrt{30}}{15}\,\lambda$ & $\dfrac{4\,\sqrt{10}}{15}\,\lambda$ & $\dfrac{4\,\sqrt{30}}{15}\,\lambda$ & $\dfrac{4\,\sqrt{30}}{15}\,\lambda$ & $\dfrac{4\,\sqrt{10}}{15}\,\lambda$ & $-\dfrac{4\,\sqrt{30}}{15}\,\lambda$ & $\dfrac{4\,\sqrt{10}}{5}\,\lambda$  & $-\dfrac{8}{15} \, \lambda^{2}$  \\[1mm]
\hline
$4$ & $\dfrac{16 \, \sqrt{10}}{9} \,\lambda$ & $0$ & $0$ & $\dfrac{16 \, \sqrt{2}}{9} \,\lambda$ & $0$ & $\dfrac{16 \, \sqrt{2}}{9} \,\lambda$ & $0$ & $\dfrac{16 \, \sqrt{2}}{9} \,\lambda$  & $-\dfrac{32}{27} \, \lambda^{2}$ \\[1mm]
\hline
\end{tabular}
}
\end{center}
\caption{{\it List of the critical points at the origin of the moduli space generated only by parity-even type IIA flux backgrounds. The quantity $\,\lambda\,$ is a free parameter setting the AdS energy scale $\,V_{0}\,$ at the solutions.}} \label{table:typeIIA_vacua_bosonic}
\end{table}

\subsubsection{The IIA landscape}

This time we perform the gauge fixing by setting $h^{+}_{0}=0$. This amounts to using the SO(2) rotating the two SL(3) factors acting on $a$ and $i$ indices. After the gauge fixing, the set of critical points includes those of the bosonic setup together with a new critical point without bosonic counterpart.

\subsubsection*{Critical points with only parity even fluxes}

Switching off the set of parity-odd flux parameters inside $F_{M\dot{\mu}}$ and $\Xi_{\alpha \beta \mu}$ recovers the maximal gauged supergravities studied in ref.\cite{Dibitetto:2012ia}. This amounts to set
\beq
g_{3} = h^{+}_{0}=0 
\hspace{10mm} \textrm{ and } \hspace{10mm}
h^{+}_{1}=g_{2}=\tilde{g}_{2}=0 \ .
\eeq

The EOMs for the scalar fields at the origin can be built using the prescription introduced in the type IIB case. The full system of quadratic flux relations can be exhaustively solved and happens to contain (up to certain sign choice multiplicities) four different solutions displayed in Table~\ref{table:typeIIA_vacua_bosonic}. These AdS$_{4}$ critical points were previously obtained in ref.~\cite{Dibitetto:2012ia} and their stability properties also discussed. These ``bosonic" solutions correspond to the first four critical points in Table~\ref{table:Landscape5Points}.

\subsubsection*{A new critical point with both parity even/odd fluxes}

Next step is to turn on the parity-odd flux parameters inside $F_{M\dot{\mu}}$ and $\Xi_{\alpha \beta \mu}$. Following the same prescription as before to obtain the EOMs, a close scrutiny of solutions to the resulting quadratic flux system can be performed. In addition to the previous solutions involving only parity-even fluxes -- and some other physically equivalent realisations thereof in terms of both parity even/odd fluxes -- we find a novel critical point without a counterpart in the purely parity-even setup. However, it is compatible with just turning on metric fluxes, Romans' mass parameter $F_{0}$ and an $F_{(6)}$ flux in analogy to solutions $2$ and $4$ in Table~\ref{table:typeIIA_vacua_bosonic}. The data for this new solution is summarised in Table~\ref{table:typeIIA_vacua_fermi}.

\begin{table}[t!]
\renewcommand{\arraystretch}{1.80}
\begin{center}
\scalebox{0.77}[0.77]{
\begin{tabular}{ | c || c | c |c | c | c | c |c | c | c || c | c | c | c | c || c | }
\hline
\textrm{\textsc{id}} & $a_{0}$ & $a_{1}$ & $a_{2}$ & $a_{3}$ & $b_{0}$ & $b_{1}$ & $c_{0}$ & $c_{1}$ & $\tilde{c}_{1}$ & $g_{3}$ & $h^{+}_{0}$ & $h^{+}_{1}$ & $g_{2}$ & $\tilde{g}_{2}$ & $V_{0}$  \\[1mm]
\hline \hline
$5$ & $\dfrac{16 \, \sqrt{10}}{9} \,\lambda$ & $0$ & $0$ & $\dfrac{16 \, \sqrt{2}}{9} \,\lambda$ & $0$ & $\dfrac{16 \, \sqrt{2}}{9} \,\lambda$ & $0$ & $0$ & $-\dfrac{16 \, \sqrt{2}}{9} \,\lambda$ & $0$ & $0$ & $0$ & $\dfrac{16 \, \sqrt{2}}{9} \,\lambda$ & $\dfrac{32 \, \sqrt{2}}{9} \,\lambda$  & $-\dfrac{32}{27} \, \lambda^{2}$ \\[1mm]
\hline
\end{tabular}
}
\end{center}
\caption{{\it New critical point at the origin of the moduli space generated by parity even/odd type IIA flux backgrounds. The quantity $\,\lambda\,$ is a free parameter setting the AdS energy scale $\,V_{0}\,$ at the solution.}} \label{table:typeIIA_vacua_fermi}
\end{table}

The mass spectrum for the vectors and scalars at this critical point can be obtained using the mass formulae (\ref{Mass-matrix_vectors}) and (\ref{Mass-matrix}). The vector masses are found to be
\beq
\label{new_spectrum_vectors}
\begin{array}{ccrrrrrrrrr}
m^2 \, L^2 &=& 15 \pm \sqrt{129} \,\,(\times 3) & , & 20 \,\,(\times 5) & , & 14 \,\,(\times 6) & , & 12 \,\,(\times 4) & , & 8 \,\,(\times 1) \\
& & 6 \,\,(\times 3) & , & 0 \,\,(\times 31) & ,
\end{array}
\eeq
whereas the masses of the scalars are given by
\beq
\label{new_spectrum_scalars}
\begin{array}{ccrrrrrrrrr}
m^2 \, L^2 &=& 21 \pm \sqrt{201} \,\,(\times 5) & , & 32 \,\,(\times 5) & , & 24 \,\,(\times 3) & , & 20 \,\,(\times 1) & , & 18 \,\,(\times 1) \\ & & 16 \,\,(\times 5) & , & 14 \,\,(\times 3) & , & 8 \,\,(\times 5) & , & 6 \,\,(\times 4) & , & 4 \,\,(\times 1) \\
& & -4 \,\,(\times 1) & , & 2 \,\,(\times 3) & , & 0 \,\,(\times 28) & .
\end{array}
\eeq
This point is non-supersymmetric, unstable with respect to scalar fluctuations  -- notice the mass eigenvalue ${m^2 L^2=-4}$ with $L^2=-3/V_{0}$ being the AdS radius -- and has an SO(3) residual symmetry reflected in the presence of $3$ massless vectors besides the $28$ unphysical ones. It corresponds with solution $5$ in Table~\ref{table:Landscape5Points}.

\subsubsection*{Discussion of the IIA landscape}

\begin{table}[t!]
\renewcommand{\arraystretch}{1.30}
\begin{center}
\scalebox{0.89}[0.89]{
\begin{tabular}{ | c || c | c |c | c | c |}
\hline
\textrm{\textsc{id}} & \textrm{\textsc{background fluxes}} & \textrm{\textsc{orientifold parity}} & \textrm{\textsc{residual sym}} & \textrm{\textsc{SUSY}}  & \textrm{\textsc{stability}}  \\[1mm]
\hline \hline
\textrm{1} & $\omega$ , $H_{(3)}$ , $F_{0}$ , $F_{(2)}$ , $F_{(4)}$ , $F_{(6)}$ & even or even/odd & SO(3) & $\mathcal{N}=1$ & $\checkmark$ \\[1mm]
\hline
\textrm{2} & $\omega$ , $F_{0}$ , $F_{(6)}$ & even or even/odd & SO(3) & $\mathcal{N}=0$ & $\times$ \\[1mm]
\hline
\textrm{3} & $\omega$ , $H_{(3)}$ , $F_{0}$ , $F_{(2)}$ , $F_{(4)}$ , $F_{(6)}$ & even or even/odd & SO(3) & $\mathcal{N}=0$ & $\checkmark$ \\[1mm]
\hline
\textrm{4} & $\omega$ , $F_{0}$ , $F_{(6)}$ & even or even/odd & SO(3)$\times$SO(3) & $\mathcal{N}=0$ & $\times$ \\[1mm]
\hline
\textrm{5} & $\omega$ , $F_{0}$ , $F_{(6)}$ & even/odd & SO(3) & $\mathcal{N}=0$ & $\times$ \\[1mm]
\hline
\end{tabular}
}
\end{center}
\caption{{\it Summary of the type IIA geometric landscape.}} 
\label{table:Landscape5Points}
\end{table}

We have summarised the results concerning the structure of the type IIA geometric landscape in Table~\ref{table:Landscape5Points}. It consists of five inequivalent critical points coexisting in a unique theory (gauging) specified by a gauge group $G=\textrm{SO}(4) \ltimes \textrm{Nil}_{22}$. This gauging was identified in ref.~\cite{Dibitetto:2012ia} and was found to be the same for the solutions $1$, $2$, $3$ and $4$ in Table~\ref{table:Landscape5Points} compatible with only parity-even fluxes. In the case of the novel solution $5$ which necessarily demands parity-odd fluxes, it can be shown that its associated flux background is connected to that of solution $4$ via a non-compact SL(2) transformation. More concretely, it acts on the indices $(a,i)$ as a doublet and maps the metric flux of solution $5$ into that of solution $4$, leaving both $F_{0}$ and $F_{(6)}$ unaffected. This mixing of $\,a\,$ and $\,i\,$ types of indices corresponds to a transformation beyond $\,\textrm{SL}(2)_{S} \times \textrm{SO}(6,6)|_{\textrm{II}}\,$ inside $\textrm{E}_{7(7)}$. In other words, the flux configuration producing the novel solution $5$ can be brought to a purely \textit{bosonic} (parity-even) one at the cost of activating some \textit{fermionic} scalars (parity-odd) which would not survive a truncation to $\mathcal{N}=4$. As a result, the new solution $5$ represents a genuine critical point of maximal supergravity which can be realised as a type IIA flux compactification on a $\mathbb{Z}_{2} \times \mathbb{Z}_{2}$ isotropic orbifold.

\begin{figure}[t!]
\begin{center}
\scalebox{0.35}[0.35]{
\includegraphics[keepaspectratio=true]{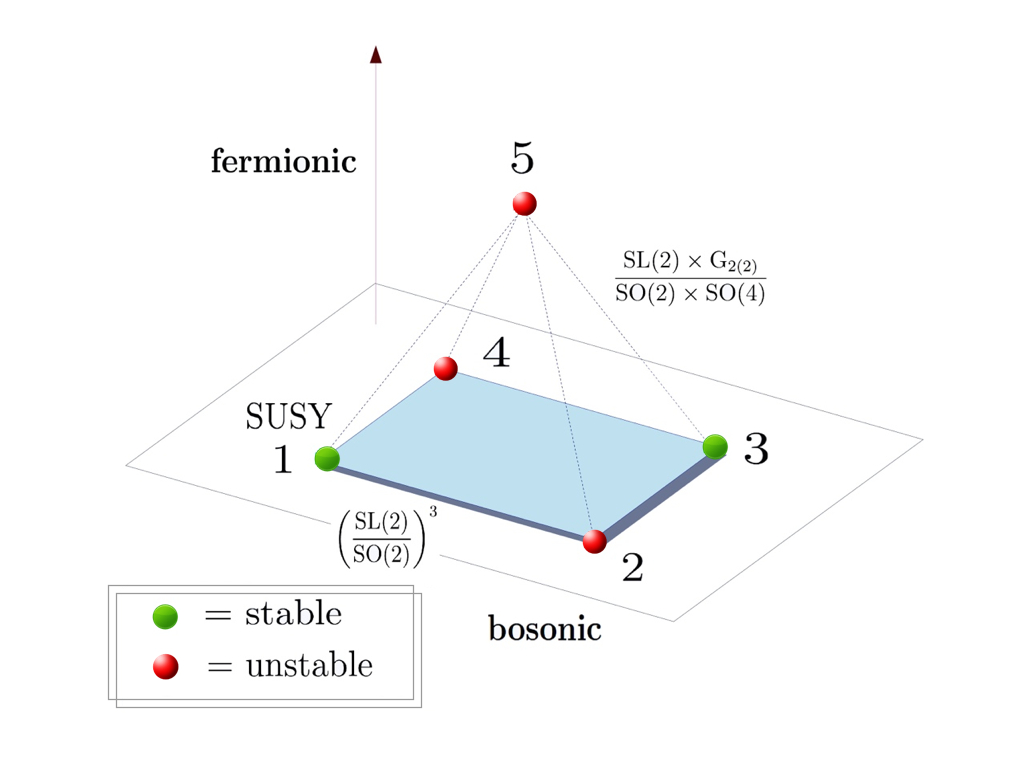}}
\end{center}
\vspace{-5mm}
\caption{{\it All the critical points of geometric type IIA compactifications (each of them represented by a vertex of the above pyramid) can be seen as different extrema of the same theory with $\textrm{SO}(4) \ltimes \textrm{Nil}_{22}$ gauge group. 
The purely bosonic solutions (labelled by 1--4), which lie on the base, have moduli vev's which are related by non-compact $\textrm{SL}(2)^{3}$ duality transformations. The new solution labelled by $5$,
instead, represents the apex of the pyramid depicted above and its moduli vev's are connected to the base via non-compact $\textrm{SL}(2) \times \textrm{G}_{2(2)}$ transformations, \emph{i.e.} U-duality 
transformations beyond S- and T-duality.}}
\label{fig:landscape}
\end{figure}

We are now interested in the twist induced by the metric flux $\omega$. It can be read off from the isometry algebra of the twisted torus $\mathbb{T}^{6}$, \textit{i.e.} $\,[ Z_{m} , Z_{n}] = {\omega_{mn}}^{p} \, Z_{p}\,$, and should match the semi-simple part of the gauge group. Using the dictionary in Table~\ref{table:IIA_ET/Fluxes}, we can rewrite the isometry brackets as
\beq
\label{isometry_algebra}
\begin{array}{ccrcrc}
\left[Z_{a} \,,\, Z_{b} \right] &=& \tilde{c}_{1}\, Z_{c} &+& h^{+}_{1} \, Z_{k} & , \\
\left[Z_{a} \,,\, Z_{j} \right] &=& g_{2}\, Z_{c} &+& c_{1} \, Z_{k} & , \\
\left[Z_{i} \,,\, Z_{j} \right] &=& b_{1} \, Z_{c} &+& \tilde{g}_{2} \, Z_{k} & ,
\end{array}
\eeq
in terms of the different components of the metric flux. The closure of this algebra is guaranteed by the Jacobi identities in (\ref{IIA_nilpotency}). An immediate way of identifying the isometry algebra in (\ref{isometry_algebra}) is to compute the associated Killing-Cartan metric $\mathcal{K}_{mn}={\omega_{mp}}^{q} \, {\omega_{nq}}^{p}\,$ \cite{Font:2008vd}. The isotropy restriction on the fluxes forces $\mathcal{K}$ to have a block-diagonal form $\mathcal{K}=\mathcal{K}_{\textrm{2} \times \textrm{2}} \, \otimes \mathbb{I}_{\textrm{3} \times \textrm{3}}$ with
\beq
\label{Killing-Cartan_2x2}
\mathcal{K}_{\textrm{2} \times \textrm{2}}  = -2 \, 
\left(
\begin{array}{cc}
\tilde{c}^{2}_{1} + 2 \, h^{+}_{1} g_{2} +c^{2}_{1} & \tilde{c}_{1} g_{2} + c_{1} g_{2} + h^{+}_{1} b_{1} + c_{1} \tilde{g}_{2} \\
\tilde{c}_{1} g_{2} + c_{1} g_{2} + h^{+}_{1} b_{1} + c_{1} \tilde{g}_{2} & \tilde{g}^{2}_{2} + 2 c_{1} b_{1} +g^{2}_{2}
\end{array}
\right) \ .
\eeq
Substituting the value of the fluxes in Tables \ref{table:typeIIA_vacua_bosonic} and \ref{table:typeIIA_vacua_fermi} into (\ref{Killing-Cartan_2x2}) one finds that $\mathcal{K}_{\textrm{2} \times \textrm{2}}$ always has two negative eigenvalues upon diagonalisation. Therefore, the Killing-Cartan metric $\mathcal{K}_{mn}$ comes out with two triplets of negative eigenvalues and the corresponding twist algebra is identified with $G_{\omega\textrm{-twist}}=\textrm{SO}(3)_{a} \times \textrm{SO}(3)_{i}\,$. 

The identification of the twist group completes our exhaustive analysis of isotropic geometric type IIA flux compactifications in the absence of D6/O6 sources \cite{oai:arXiv.org:hep-th/0403049,oai:arXiv.org:hep-th/0407263,oai:arXiv.org:hep-th/0412250,oai:arXiv.org:0712.1021}. In addition to the $\mathcal{N}=1$ solution in Table~\ref{table:Landscape5Points} (solution $1$), there is a non-supersymmetric and nevertheless fully stable solution (solution $3$) requiring all types of IIA fluxes. Lastly, despite the fact that they are unstable, we want to highlight the presence of three non-supersymmetric critical points (solutions $2$, $4$ and $5$) compatible with the very strong restriction $F_{(2)}=F_{(4)}=0$, thus enjoing a fairly simple lift to \textit{massive} IIA supergravity in ten dimension. The geometric IIA landscape is depicted in Figure~\ref{fig:landscape}.

\section{Summary and final remarks}
\label{sec:conclusions}

In this work we have studied flux compactifications of type II string theories on a twisted six-torus in the absence of localised sources, \textit{i.e.} D-branes and O-planes. To this end, we have made use of their description in terms of maximal gauged supergravities in four dimensions and have explicitly derived the embedding tensor/fluxes dictionary. 

In the first part of the paper, we exploited the group theory structure underlying the embedding tensor formalism. We adopted an intermediate approach between the one proposed in ref.~\cite{Aldazabal:2010ef}, which is inspired by Exceptional Generalised Geometry, and the one recently proposed in refs~\cite{Godazgar:2013dma,Godazgar:2013pfa,Godazgar:2013oba,Godazgar:2014sla} based on generalised twisted reductions of M-theory including both the $A_{(3)}$ and $A_{(6)}$ gauge potentials. In this way, we found perfect agreement (at least at the group theory level) between 4D supergravity states and states coming from the reduction of the democratic formulation of type II supergravities \cite{Bergshoeff:2001pv} before imposing any self-duality or physical section condition on the degrees of freedom. The question of how to impose such a section condition to remove non-dynamical states and whether it would kill any orbit of truly non-geometric backgrounds deserves further investigation. Also possible links to the weak/strong constraints in the framework of (E)DFT (see \emph{e.g.} refs~\cite{Grana:2012rr,Dibitetto:2012rk,Aldazabal:2013mya,Betz:2014aia} for recent developments in this direction). We hope to come back to these and other related issues in the future.

In the second part of the paper, we tested the embedding tensor/fluxes dictionary. We explored the most general geometric flux backgrounds of both type IIA and type IIB strings on an isotropic $\mathbb{T}^{6}/(\mathbb{Z}_{2} \times \mathbb{Z}_{2})$ orbifold and exhaustively analysed their vacuum structure. Surprisingly, within this class of theories, there turns out to be a unique flux compactification allowing for critical points, corresponding to an ${\textrm{SO}(4) \ltimes \textrm{Nil}_{22}}$ gauging. Beyond the four AdS critical points already found in ref.~\cite{Dibitetto:2011gm} and further investigated in ref.~\cite{Dibitetto:2012ia}, which admitted a truncation to half-maximal supergravity, a new AdS solution is found, which occurs at non-vanishing vev's for scalars beyond the $\frac{\textrm{SL}(2)}{\textrm{SO}(2)} \times \frac{\textrm{SO}(6,6)}{\textrm{SO}(6)\times \textrm{SO}(6)}$ coset spanned by the $\cN=4$ scalars. From a supergravity viewpoint, the above new solution, which then turns out to be non-supersymmetric and unstable, provides us with a novel example of a critical point of maximal supergravity with $\textrm{SO}(3)$ residual symmetry. It exhibits a new mass spectrum, which might then hint at possible holographic applications in the context of the gauge/gravity correspondence. From a stringy perspective, this set of five AdS solutions provides an exhaustive classification of isotropic extrema of type II strings compactified on $\mathbb{T}^{6}/(\mathbb{Z}_{2} \times \mathbb{Z}_{2})$ in the absence of localised sources. 

Geometric compactifications are generically compatible with a large volume and small string coupling regime where all corrections can be kept under perturbative control.  However, when trying to perform this in practice, one realises that it is done through a scaling of flux quanta to very large values, which has the desirable feature of hiding flux quantisation, but at the same time it generates an inconsistency with the cancellation of the O-plane charge. Due to the absence of O-planes and D-branes, such exceptional string backgrounds offer the possibility to achieve all
of this without encountering the above problem.

%
%

\section*{Acknowledgments}

The work of GD is supported by the Swedish Research Council (VR), and the G\"oran Gustafsson Foundation. The work of AG is supported by the Swiss National Science Foundation. The work of DR is supported by a VIDI grant from the Netherlands Organisation for Scientific Research (NWO). AG wants to thank the hospitality of the Department of Physics and Astronomy at Uppsala University where part of this work was completed.

%
%

\newpage
\appendix

\section{The mapping between polyforms and spinors}
\label{App:polyforms}

In this appendix we discuss in detail how does the correspondence between M-W spinors of $\,\textrm{SO}(6,6)\,$ and polyforms of $\,\textrm{SL}(6)\,$ work. This correspondence became of utmost importance
in Section~\ref{sec:Fluxes/ET} where the embedding of $\,\textrm{SL}(6)\,$ fluxes into $\,\textrm{SO}(6,6)\,$ M-W spinors was extensively used.
Given a left-handed M-W spinor $\,T_{\mu}\,$, it can always be mapped into a sum of antisymmetric $\,p$-forms of odd degree $\,p=1,3,5\,$, namely,
\beq
\label{polyform_LH}
T_{\mu} \,\, = \,\, T_{\underline{m}} \,\, \oplus \,\, T_{[\underline{m_{1}...m_{3}}]} \,\, \oplus \,\, T_{[\underline{m_{1}...m_{5}}]} \,\, = \,\, T_{\underline{m}} \,\, \oplus \,\, T_{[\underline{m_{1}...m_{3}}]} \,\, \oplus \,\, T^{\underline{m}} \ ,
\eeq
where $\,T^{\underline{m}}= \frac{1}{5!} \, \epsilon^{\underline{m n_{1}...n_{5}}} \, T_{[\underline{n_{1}...n_{5}}]}\,$. Analogously, provided a right-handed M-W spinor $\,T_{\dot{\m}}\,$, it can be decomposed into a sum of antisymmetric $\,p$-forms of even degree $\,p=6,2,4,0\,$. This is
\beq
\begin{array}{ccl}
\label{polyform_RH}
T_{\dot{\mu}} & = & T_{[\underline{m_{1}...m_{6}}]} \,\, \oplus \,\, T_{[\underline{m_{1}m_{2}}]} \,\, \oplus \,\, T_{[\underline{m_{1}...m_{4}}]} \,\, \oplus \,\, T \\[3mm]
& = & T_{[\underline{m_{1}...m_{6}}]} \,\, \oplus \,\, T_{[\underline{m_{1}m_{2}}]} \,\, \oplus \,\, T^{[\underline{m_{1}m_{2}}]} \,\, \oplus \,\, T^{[\underline{m_{1}...m_{6}}]} \ , \\[2mm]
\end{array}
\eeq
with $\,T^{[\underline{m_{1}m_{2}}]}= \frac{1}{4!} \, \epsilon^{\underline{m_{1}...m_{6}}} \, T_{[\underline{m_{3}...m_{6}}]}\,\,$ and $\,\,T^{[\underline{m_{1}...m_{6}}]}= \epsilon^{\underline{m_{1}...m_{6}}} \, T\,$. In the following, we will make the above spinor/polyforms correspondences more precise.
Let us start by introducing a set of $\,8 \times 8\,$ matrices $\,\left\lbrace \Sigma_{\underline{m}} \right\rbrace_{\underline{m}=1,...,6}\,$ spanning a time-like $\,\textrm{SO}(6)\,$ Clifford algebra in the Dirac representation
\beq
\label{Dirac_algebra}
\left\lbrace \Sigma_{\underline{m}} \, , \, \Sigma_{\underline{n}}\right\rbrace = - \, 2 \, \delta_{\underline{mn}} \, \mathds{I}_{8}\ .
\eeq
We adopt the conventions in which an $\,\textrm{SO}(6)\,$ Dirac spinor carries an upper index $\,\psi^{I}\,$, with $\,I=1,...,8\,$, so the $\Sigma_{\underline{m}}$-matrices come out with an index structure $\,{[\Sigma_{\underline{m}}]^{I}}_{J}\,$ to properly act upon it. Moving to a Weyl basis for the algebra (\ref{Dirac_algebra}), a Dirac spinor splits into left- and right-handed components $\,\psi^{I} = \left( \, \psi^{i} \, , \, \psi_{\hat{i}} \, \right)\,$, with $\,i,\hat{i}=1,...,4\,$, and the set of $\,\Sigma_{\underline{m}}\,$ matrices take the off-block-diagonal form
\beq
{[\Sigma_{\underline{m}}]^{I}}_{J}
=
\left(
\begin{array}{cc}
0 & [\sigma_{\underline{m}}]^{i \hat{j}} \\
{[\bar{\sigma}_{\underline{m}}]}_{\hat{i} j} & 0
\end{array}
\right) \ .
\eeq
The Dirac charge conjugation matrix $\,C \equiv C_{IJ}\,$ entering the relations $\,\Sigma_{\underline{m}}^{T}=-C \, \Sigma_{\underline{m}} \, C^{-1}\,$ takes the form
\beq
\label{charge_conjug}
C_{IJ}
=
\left(
\begin{array}{cc}
0 & {C_{i}}^{\hat{j}} = i \, \eta_{13} \\
{\bar{C}^{\hat{i}}}_{\,\,j}=i \, \eta_{13} & 0
\end{array}
\right)
\eeq
where $\,\eta_{13} = \textrm{diag}(-1,1,1,1)\,$ and moreover $\,C^{*} = C^{-1} \equiv C^{IJ}\,$. The charge conjugation matrix in (\ref{charge_conjug}) is compatible with taking the following set of $\,[\sigma_{\underline{m}}]^{i \hat{j}}\,$ matrices
\begin{equation}
\label{sigma_matrices}
\begin{array}{c}
[\s_{1}]=
{\scriptsize{\left[
\begin{array}{cccc}
0 & 1 & 0 & 0 \\
1 & 0 & 0 & 0 \\
0 & 0 & 0 & -1 \\
0 & 0 & 1 & 0
\end{array}
\right]}}
\hspace{2mm} , \hspace{2mm}
[\s_{3}]=
{\scriptsize{\left[
\begin{array}{cccc}
0 & 0 & 1 & 0 \\
0 & 0 & 0 & 1 \\
1 & 0 & 0 & 0 \\
0 & -1 & 0 & 0
\end{array}
\right]}}
\hspace{2mm} , \hspace{2mm}
[\s_{5}]=
{\scriptsize{\left[
\begin{array}{cccc}
0 & 0 & 0 & 1 \\
0 & 0 & -1 & 0 \\
0 & 1 & 0 & 0 \\
1 & 0 & 0 & 0
\end{array}
\right]}} \ ,
\\[8mm]
[\s_{2}]=
{\scriptsize{\left[
\begin{array}{cccc}
0 & i & 0 & 0 \\
i & 0 & 0 & 0 \\
0 & 0 & 0 & i \\
0 & 0 & -i & 0
\end{array}
\right]}}
\hspace{2mm} , \hspace{2mm}
[\s_{4}]=
{\scriptsize{\left[
\begin{array}{cccc}
0 & 0 & i & 0 \\
0 & 0 & 0 & -i \\
i & 0 & 0 & 0 \\
0 & i & 0 & 0
\end{array}
\right]}}
\hspace{2mm} , \hspace{2mm}
[\s_{6}]=
{\scriptsize{\left[
\begin{array}{cccc}
0 & 0 & 0 & i \\
0 & 0 & i & 0 \\
0 & -i & 0 & 0 \\
i & 0 & 0 & 0
\end{array}
\right]}} \ ,
\end{array}
\end{equation}
together with the $\,[\bar{\sigma}_{\underline{m}}]_{\hat{i} j}\,$ ones
\begin{equation}
\label{sigma_bar_matrices}
\begin{array}{c}
[\bar{\s}_{1}]=
{\scriptsize{\left[
\begin{array}{cccc}
0 & -1 & 0 & 0 \\
-1 & 0 & 0 & 0 \\
0 & 0 & 0 & -1 \\
0 & 0 & 1 & 0
\end{array}
\right]}}
\hspace{2mm} , \hspace{2mm}
[\bar{\s}_{3}]=
{\scriptsize{\left[
\begin{array}{cccc}
0 & 0 & -1 & 0 \\
0 & 0 & 0 & 1 \\
-1 & 0 & 0 & 0 \\
0 & -1 & 0 & 0
\end{array}
\right]}}
\hspace{2mm} , \hspace{2mm}
[\bar{\s}_{5}]=
{\scriptsize{\left[
\begin{array}{cccc}
0 & 0 & 0 & -1 \\
0 & 0 & -1 & 0 \\
0 & 1 & 0 & 0 \\
-1 & 0 & 0 & 0
\end{array}
\right]}} \ ,
\\[8mm]
[\bar{\s}_{2}]=
{\scriptsize{\left[
\begin{array}{cccc}
0 & i & 0 & 0 \\
i & 0 & 0 & 0 \\
0 & 0 & 0 & -i \\
0 & 0 & i & 0
\end{array}
\right]}}
\hspace{2mm} , \hspace{2mm}
[\bar{\s}_{4}]=
{\scriptsize{\left[
\begin{array}{cccc}
0 & 0 & i & 0 \\
0 & 0 & 0 & i \\
i & 0 & 0 & 0 \\
0 & -i & 0 & 0
\end{array}
\right]}}
\hspace{2mm} , \hspace{2mm}
[\bar{\s}_{6}]=
{\scriptsize{\left[
\begin{array}{cccc}
0 & 0 & 0 & i \\
0 & 0 & -i & 0 \\
0 & i & 0 & 0 \\
i & 0 & 0 & 0
\end{array}
\right]}} \ .
\end{array}
\end{equation}
With the above sets (\ref{sigma_matrices}) and (\ref{sigma_bar_matrices}) of $\,\sigma_{\underline{m}}\,$ and $\,\bar{\sigma}_{\underline{m}}\,$ matrices we can go further and also build complete sets of $\,\s_{(p)}$-forms up to $\,p=6\,$. In the case of even values of $\,p=0,2,4,6\,$, one obtains
\beq
\hspace{-1.7mm}
\begin{array}{rclc}
{[\s_{(0)}]_{i}}^{\hat{j}} & = & {C_{i}}^{\hat{j}} & ,\\[2mm]
{[\s_{\underline{m_{1}m_{2}}}]_{i}}^{\hat{j}} & = & {C_{i}}^{\hat{k}_{1}} \, [\bar{\s}_{[\underline{m_{1}}}]_{\hat{k}_{1} k_{2}} \, [\s_{\underline{m_{2}}]}]^{k_{2}\hat{j}} & ,\\[2mm]
{[\s_{\underline{m_{1}m_{2}m_{3}m_{4}}}]_{i}}^{\hat{j}} & = & {C_{i}}^{\hat{k}_{1}} \, [\bar{\s}_{[\underline{m_{1}}}]_{\hat{k}_{1} k_{2}} \, [\s_{\underline{m_{2}}}]^{k_{2}\hat{k}_{3}} \, [\bar{\s}_{\underline{m_{3}}}]_{\hat{k}_{3} k_{4}} \, [\s_{\underline{m_{4}}]}]^{k_{4}\hat{j}}& , \\[2mm]
{[\s_{\underline{m_{1}m_{2}m_{3}m_{4}m_{5}m_{6}}}]_{i}}^{\hat{j}} & = & {C_{i}}^{\hat{k}_{1}} \, [\bar{\s}_{[\underline{m_{1}}}]_{\hat{k}_{1} k_{2}} \, [\s_{\underline{m_{2}}}]^{k_{2}\hat{k}_{3}} \, [\bar{\s}_{\underline{m_{3}}}]_{\hat{k}_{3} k_{4}} \, [\s_{\underline{m_{4}}}]^{k_{4}\hat{k}_{5}} \, [\bar{\s}_{\underline{m_{5}}}]_{\hat{k}_{5} k_{6}} \, [\s_{\underline{m_{6}}]}]^{k_{6}\hat{j}} & ,
\end{array}
\eeq
together with their complex conjugates $\,{[\s_{(p)}]^{i}}_{\hat{j}} = \left( {[\s_{(p)}]_{i}}^{\hat{j}} \right)^{*}\,$. Equivalently, for odd values of $\,p=1,3,5\,$, one finds
\beq
\begin{array}{rclc}
[\s_{\underline{m}}]^{i\hat{j}} & = & [\s_{\underline{m}}]^{i\hat{j}} & ,\\[2mm]
[\s_{\underline{m_{1}m_{2}m_{3}}}]^{i \hat{j}} & = & [\s_{[\underline{m_{1}}}]^{i\hat{k}_{1}} \, [\bar{\s}_{\underline{m_{2}}}]_{\hat{k}_{1} k_{2}} \, [\s_{\underline{m_{3}}]}]^{k_{2}\hat{j}} & ,\\[2mm]
[\s_{\underline{m_{1}m_{2}m_{3}m_{4}m_{5}}}]^{i\hat{j}} & = & [\s_{[\underline{m_{1}}}]^{i\hat{k}_{1}} \, [\bar{\s}_{\underline{m_{2}}}]_{\hat{k}_{1} k_{2}} \, [\s_{\underline{m_{3}}}]^{k_{2}\hat{k}_{3}} \, [\bar{\s}_{\underline{m_{4}}}]_{\hat{k}_{3} k_{4}} \, [\s_{\underline{m_{5}}]}]^{k_{4}\hat{j}} & ,
\end{array}
\eeq
and, once again, there are also their complex conjugates $\,[\s_{(p)}]_{i\hat{j}} = \left( [\s_{(p)}]^{i\hat{j}} \right)^{*}\,$. In order to derive the spinor/polyforms mapping, we will make use of the counterparts of the previous $\s_{(p)}$-forms with upper indices. They are defined as
\beq
\label{sigma_upper}
[\sigma^{\underline{m_{1}...m_{p}}}] = \frac{1}{(6-p)!} \, \epsilon^{\underline{m_{1}...m_{6}}} \, [\sigma_{\underline{m_{p+1}...m_{6}}}] \hspace{10mm} \textrm{ for } \hspace{5mm} p=0,...,6 \ .
\eeq
The precise spinor/polyforms correspondence can now be introduced. As a preliminary step, we must decompose $\,\textrm{SO(6,6)}\,$ M-W spinors with respect to its $\,\textrm{SO}(6) \times \textrm{SO}(6) \sim {\textrm{SU}(4) \times \textrm{SU}(4)} \,$ maximal subgroup. This produces the branchings $\,\textbf{32} \rightarrow (\textbf{4},\textbf{4}) \,+\, (\bar{\textbf{4}},\bar{\textbf{4}})\,$ and $\,\textbf{32'} \rightarrow (\textbf{4},\bar{\textbf{4}}) \,+\, (\bar{\textbf{4}},\textbf{4})\,$, and amounts to the decompositions
\beq
\label{32_decomp}
T_{\mu} \,\,=\,\, T_{i \hat{j}} \,\, \oplus \,\, T^{i \hat{j}}
\hspace{10mm} \textrm{ and } \hspace{10mm}
T_{\dot{\mu}} \,\,=\,\, {T_{i}}^{\hat{j}} \,\, \oplus \,\, {T^{i}}_{\hat{j}} \ .
\eeq
Considering a diagonal $\,\textrm{SU}(4)_{\textrm{D}}\,$ subgroup (in order to deal with bi-spinors) and using the $\s^{(p)}$-forms in (\ref{sigma_upper}), the final mapping is given by
\beqa
\label{polyform_final}
T_{i \hat{j}} &=& T_{\underline{m}} \, [\sigma^{\underline{m}}]_{i \hat{j}} \,+\, \frac{1}{3!} \, T_{[\underline{m_{1}...m_{3}}]} \, [\sigma^{\underline{m_{1}...m_{3}}}]_{i \hat{j}} \,+\, \frac{1}{5!} \, T_{[\underline{m_{1}...m_{5}}]} \, [\sigma^{\underline{m_{1}...m_{5}}}]_{i \hat{j}} \ , \nonumber \\[-4mm]
& & \\[0mm]
{T_{i}}^{\hat{j}} &=& \frac{1}{6!} \, T_{[\underline{m_{1}...m_{6}}]} \, {[\sigma^{\underline{m_{1}...m_{6}}}]_{i}}^{\hat{j}} \,+\, \frac{1}{2!} \, T_{[\underline{m_{1}m_{2}}]} \, {[\sigma^{\underline{m_{1}m_{2}}}]_{i}}^{\hat{j}} \,+\, \frac{1}{4!} \, T_{[\underline{m_{1}...m_{4}}]} \, {[\sigma^{\underline{m_{1}...m_{4}}}]_{i}}^{\hat{j}} \,+ \,T \, {[\sigma^{(0)}]_{i}}^{\hat{j}} \nonumber \ ,
\eeqa
together with their complex conjugates $\,T^{i \hat{j}}=(T_{i \hat{j}})^{*}\,$ and $\,{T^{i}}_{\hat{j}}=({T_{i}}^{\hat{j}})^{*}\,$. The terms $\,T_{[\underline{m_{1}...m_{p}}]}\,$ with $\,p=0,...,6\,$ in the r.h.s of (\ref{polyform_final}) are in one-to-one correspondence with those in (\ref{polyform_LH}) and (\ref{polyform_RH}). Then, by using (\ref{polyform_final}) and subsequently (\ref{32_decomp}) one obtains the spinor/polyforms mapping
\beq
\label{mapping_final}
\begin{array}{lccccccc}
p= \textrm{odd} & : \hspace{10mm} & \displaystyle{\bigoplus_{p=0}^{6}} \, T_{[\underline{m_{1}...m_{p}}]} & \longrightarrow & \big \lbrace \, T_{i \hat{j}} \,\, , \,\, T^{i \hat{j}} \, \big \rbrace & \longrightarrow & T_{\mu} & , \\[2mm]
p= \textrm{even} & : \hspace{10mm} & \displaystyle{\bigoplus_{p=0}^{6}} \, T_{[\underline{m_{1}...m_{p}}]} & \longrightarrow & \big \lbrace \, {T_{i}}^{\hat{j}} \,\, , \,\, {T^{i}}_{\hat{j}} \, \big \rbrace & \longrightarrow & T_{\dot{\mu}} & ,
\end{array}
\eeq
for left- and right-handed M-W spinors of $\,\textrm{SO}(6,6)$, respectively. This mapping plays a central role in deriving the complete embedding tensor/fluxes dictionary including also the orientifold-odd components.

\newpage

\section{Dimensional reductions of type II string theory}
\label{App:Dim_Red}

In this appendix we discuss some conventions related to dimensional reductions of type II string theory on a $\mathbb{T}^{6}$ down to four dimensions.
The low-energy type IIB (pseudo-)action in the string frame reads
\bea
S^{(\textrm{IIB})} & = & \dfrac{1}{(2\pi)^{7}\,(\a^{\prime})^{4}} \, \int d^{10}x \, \sqrt{-g_{10}} \, \left(e^{-2\phi}\mathcal{R}^{(10)} \, + \, 4 e^{-2\phi} (\partial\phi)^{2}
\, - \, \dfrac{1}{2 \cdot 3!}e^{-2\phi} |H_{3}|^{2}\right. \, + \notag \\
& & \left. - \,\dfrac{1}{2} |F_{1}|^{2}\, - \, \dfrac{1}{2 \cdot 3!} |F_{3}|^{2}\, - \, \dfrac{1}{2 \cdot 5!} |F_{5}|^{2}\right) \, + \, \textrm{C-S} \ , \label{IIB_action}
\eea
where $F_{5}$ should satisfy $F_{5} \, \overset{!}{=} \, \star_{10}F_{5}$.
We choose the following reduction \emph{Ansatz}
\be
\label{IIB_red_ansatz}
ds^{2}_{10} \, = \, \tau^{-2} \, ds^{2}_{4} \, + \, \rho \, M_{mn} \, dy^{m} \, dy^{n} \ ,
\ee
where $\tau$ and $\rho$ are suitable combinations of the internal volume $\textrm{vol}_{6}$ and the ten-dimensional dilaton $\phi$ which are usually referred to as the universal moduli
\cite{Hertzberg:2007wc}. The internal geometry is parametrised by the element $M_{mn}$ of the $\textrm{SL}(6)/\textrm{SO}(6)$ coset.
According to \eqref{IIB_red_ansatz}, the ten-dimensional Ricci scalar $\mathcal{R}^{(10)}$ reduces to
\be
\label{Ricci_red}
\begin{array}{cccc}
\mathcal{R}^{(10)} & \longrightarrow & \tau^{2} \, \mathcal{R}^{(4)} \, + \, \rho^{-1} \, \mathcal{R}^{(6)} & .
\end{array}
\ee
Imposing
\be
\label{Eistein_4D}
e^{2\phi} \, = \, \tau^{-2}\rho^{3}
\ee
guarantees a four-dimensional Lagrangian in the Einstein frame. By performing the dimensional reduction of the various kinetic terms in the action \eqref{IIB_action}, one can derive the $(\rho,\tau)$ scaling of the corresponding fluxes in a very
straightforward way. Subsequently, by observing that these scalars are related to the dilatons sitting inside $S$ and $T$ in the following way
\be
\label{rhotau/ST}
\rho  =  \textrm{Im}(S)^{-1/2} \, \textrm{Im}(T)^{1/2}
\hspace{10mm} , \hspace{10mm} 
\tau = \textrm{Im}(S)^{1/4} \, \textrm{Im}(T)^{3/4} \ ,
\ee
one can read off their $ST$ weights as given in Section~\ref{sec:GroupI_embedding_IIA}. As an example, let us derive the $ST$ weights of $F_{mnp}$. By dimensional reduction according to \eqref{IIB_red_ansatz}, one finds
\be
\begin{array}{lclclc}
\sqrt{-g_{10}} \, |F_{3}|^{2} & \longrightarrow & \tau^{-4}\rho^{3} \, |F_{mnp}|^{2} \, \rho^{-3} & = &
\tau^{-4} \, |F_{mnp}|^{2} & ,
\end{array}
\ee
where $|F_{mnp}|^{2} \, \equiv \, F_{mnp}F_{qrs}M^{mq}M^{nr}M^{ps}$. Using the invariance of the scalar potential together with the mapping \eqref{rhotau/ST}, one finds
\be
\begin{array}{lclclc}
F_{mnp} & \sim & \textrm{Im}(S)^{1/2} \, \textrm{Im}(T)^{3/2} & ,
\end{array}
\ee
which is in perfect agreement with the $ST$ weights given in the second row of Table~\ref{table:IIB_Fluxes}.

The low-energy \textit{massive} type IIA action in the string frame reads
\bea
S^{(\textrm{IIA})} & = & \dfrac{1}{(2\pi)^{7}\,(\a^{\prime})^{4}} \, \int d^{10}x \, \sqrt{-g_{10}} \, \left(e^{-2\phi}\mathcal{R}^{(10)} \, + \, 4 e^{-2\phi} (\partial\phi)^{2}
\, - \, \dfrac{1}{2 \cdot 3!}e^{-2\phi} |H_{3}|^{2}\right. \, + \notag \\
& & \left. - \,\dfrac{1}{2} |F_{0}|^{2}\, - \, \dfrac{1}{2 \cdot 2!} |F_{2}|^{2}\, - \, \dfrac{1}{2 \cdot 4!} |F_{4}|^{2}
\, - \, \dfrac{1}{2 \cdot 6!} |F_{6}|^{2}\right) \, + \, \textrm{C-S} \ . 
\label{IIA_action}
\eea
This time we choose the following reduction \emph{Ansatz}
\be
\label{IIA_red_ansatz}
ds^{2}_{10} \, = \, \tau^{-2} \, ds^{2}_{4} \, + \, \rho \, \left(\sigma^{-3} \, M_{ab} \, dy^{a} \, dy^{b} \, + \, \sigma^{3} \, M_{ij} \, dy^{i} \, dy^{j}\right) \ ,
\ee
where $\tau$ and $\rho$ are defined as in \eqref{IIB_red_ansatz}. The extra $\mathbb{R}^{+}$ scalar $\sigma$ parametrises the relative size between the $a$ and $i$ coordinates \cite{Danielsson:2012et}, whereas $M_{ab}$ and $M_{ij}$
contain $\textrm{SL}(3)_{a} \, \times \, \textrm{SL}(3)_{i}$ scalars. As a consequence of \eqref{IIA_red_ansatz}, the ten-dimensional Ricci scalar $\mathcal{R}^{(10)}$ still reduces according to the universal form described in \eqref{Ricci_red} for the type IIB case. Moreover, imposing \eqref{Eistein_4D} still gives a four-dimensional Lagrangian in the Einstein frame.
By performing the dimensional reduction of the various kinetic terms in the action \eqref{IIA_action}, one can derive the $(\rho,\tau,\sigma)$ scaling of the corresponding fluxes. Using the relation between these scalars and the dilatons sitting inside $S$, $T$ and $U$ given by
\be
\label{rhotausigma/STU}
\rho  =  \textrm{Im}(U) 
\hspace{5mm} , \hspace{5mm}
\tau  =  \textrm{Im}(S)^{1/4} \, \textrm{Im}(T)^{3/4} 
\hspace{5mm} , \hspace{5mm}
\sigma  =  \textrm{Im}(S)^{-1/6} \, \textrm{Im}(T)^{1/6} \ ,
\ee
one can read off their $STU$ weights as given in Section~\ref{sec:GroupI_embedding_IIA}. To illustrate this, let us derive the $STU$ weights of $H_{ijk}$. By dimensional reduction according to \eqref{IIA_red_ansatz}, this time one finds
\be
\begin{array}{lclclc}
\sqrt{-g_{10}} \, e^{-2\phi} |H_{3}|^{2} & \longrightarrow & \tau^{-4}\rho^{3} \, e^{-2\phi} \, |H_{ijk}|^{2} \, \sigma^{-9}\rho^{-3} & \overset{\eqref{Eistein_4D}}{=} &
\rho^{-3}\tau^{-2}\sigma^{-9} \, |H_{ijk}|^{2} & ,
\end{array}
\ee
where $|H_{ijk}|^{2} \, \equiv \, H_{ijk}H_{i'j'k'}M^{ii'}M^{jj'}M^{kk'}$. Using again the invariance of the scalar potential together with the mapping \eqref{rhotausigma/STU}, one gets
\be
\begin{array}{lclclc}
H_{ijk} & \sim & \textrm{Im}(S)^{-1/2} \, \textrm{Im}(T)^{3/2} \, \textrm{Im}(U)^{3/2} & ,
\end{array}
\ee
which is also in perfect agreement with the $STU$ weights given in the fourth row of Table~\ref{table:IIA_Fluxes}. Finally, because of its relevance in Section~\ref{sec:GroupI_embedding_IIA}, we will compute the STU weights of the Roman's mass $F_{0}$. Upon dimensional reduction, the relevant term in the action (\ref{IIA_action}) reads
\be
\begin{array}{lclclc}
\sqrt{-g_{10}} \, |F_{0}|^{2} & \longrightarrow & \tau^{-4}\rho^{3} \, |F_{0}|^{2}  \  .
\end{array}
\ee
Using again the identifications in (\ref{rhotausigma/STU}), the invariance of the scalar potential demands
\beq
\begin{array}{lclclc}
F_{0} & \sim & \textrm{Im}(S)^{1/2} \, \textrm{Im}(T)^{3/2} \, \textrm{Im}(U)^{-3/2} & ,
\end{array}
\eeq
hence being in perfect agreement with the result in Table~\ref{table:IIA_Fluxes}.

\newpage

%
%
\small
\bibliography{references}
\bibliographystyle{utphys}
\end{document}